\documentclass[pr,superscriptaddress,twocolumn,amsmath]{revtex4-1}

\usepackage{bm}
\usepackage{graphicx}
\usepackage[usenames,dvipsnames,svgnames,table]{xcolor}
\usepackage[colorlinks=true,linkcolor=RoyalBlue,citecolor=RoyalBlue]{hyperref}

\begin{document}

\title{Magneto-optical Kerr and Faraday effects in bilayer antiferromagnetic insulators}

\author{Wan-Qing Zhu}
\affiliation{Department of Physics, School of Physics and Materials Science, Guangzhou University, Guangzhou 510006, China}
\author{Wen-Yu Shan}
\email{wyshan@gzhu.edu.cn}
\affiliation{Department of Physics, School of Physics and Materials Science, Guangzhou University, Guangzhou 510006, China}

\date{\today}

\begin{abstract}
Control and detection of antiferromagnetic topological materials are challenging since the total magnetization vanishes. Here we investigate the magneto-optical Kerr and Faraday effects in bilayer antiferromagnetic insulator MnBi$_2$Te$_4$.  We find that by breaking the combined mirror symmetries with either perpendicular electric field or external magnetic moment, Kerr and Faraday effects occur. Under perpendicular electric field, antiferromagnetic topological insulators (AFMTI) show sharp peaks at the interband transition threshold, whereas trivial insulators show small adjacent positive and negative peaks. Gate voltage and Fermi energy can be tuned to reveal the differences between AFMTI and trivial insulators. We find that AFMTI with large antiferromagnetic order can be proposed as a pure magneto-optical rotator due to sizable Kerr (Faraday) angles and vanishing ellipticity. Under external magnetic moment, AFMTI and trivial insulators are significantly different in the magnitude of Kerr and Faraday angles and ellipticity. For the qualitative behaviors, AFMTI shows distinct features of Kerr and Faraday angles when the spin configurations of the system change. These phenomena provide new possibilities to optically detect and manipulate the layered topological antiferromagnets.

\end{abstract}

\maketitle

\section{Introduction}

Recent years have seen a surge of interest in antiferromagnetic (AFM) materials due to their advantages in spintronics applications: they are robust against perturbation due to magnetic field, display ultrafast dynamics,  and so on~\cite{jungwirth2016,baltz2018}. When combining AFM with topology, new topological phenomena arise, including the condensed matter version of axions~\cite{li2010,nenno2020,liu2020}, quantized magnetoelectric couplings~\cite{essin2009,zhang2019,armitage2019} and anomalous Hall effect~\cite{chen2014,nakatsuji2015,nayak2016}. These topological effects largely depend on the spin structures of AFM systems. For example, in noncollinear triangular AFM, anomalous Hall effect occurs due to the chirality of spin structures~\cite{chen2014,nakatsuji2015,nayak2016}. In noncoplanar AFM, quantized magneto-optical responses arise from the finite scalar spin chirality~\cite{feng2020}.

An intriguing example is the van der Waals compound MnBi$_2$Te$_4$, which, as an A-type antiferromagnetic topological insulator (AFMTI)~\cite{otrokov2019,otrokovm2019,chen2019,hao2019,swatek2020}, shows anti-parallel magnetization between adjacent layers and zero total magnetization. Due to the combined $\mathcal{PT}$ symmetry ($\mathcal{P}$ and $\mathcal{T}$ stand for inversion and time-reversal symmetry), even-layered  MnBi$_2$Te$_4$ is reported to exhibit large longitudinal resistance with zero Hall plateau~\cite{liu2020}. Accordingly, the even-layered system is predicted to host the exotic axion insulator phase~\cite{zhang2019,liu2020,zhang2020}. One evidence distinguishing AFMTI, i.e., axion insulator, from trivial insulator is the layer Hall effect~\cite{gao2021,chen2022}, where electrons from top and bottom layers move in opposite transverse directions due to the layer-locked Berry curvature. Other proposals based on different transport phenomena are also raised to identify the axion state~\cite{chen2021,ding2020,zhangr2020,li2021,gu2021,dai2022}. By contrast, optical probe of AFMTI or axion insulator has rarely been discussed~\cite{ahn2022,lei2023,qiu2023}. Particularly, for few even-layer MnBi$_2$Te$_4$, whether AFMTI and trivial phases are distinguishable in optical responses remains elusive.

In this work, we study the magneto-optical Kerr and Faraday effects in bilayer AFM insulators MnBi$_2$Te$_4$. We find that to allow non-vanishing Kerr and Faraday effects, the combined mirror $\mathcal{M}_{1}=\mathcal{M}_{x}\mathcal{M}_{z}$ and $\mathcal{M}_{2}=\mathcal{M}_{y}\mathcal{M}_{z}$ symmetries ($\mathcal{M}_{x/y/z}$ stands for mirror symmetry) should be broken by adding either perpendicular electric field or external magnetic moment (see Fig.~\ref{fig:setup}). In the former case, we find that Kerr and Faraday angles of AFMTI have significant peaks at the transition threshold as a result of non-monotonic band dispersion, in contrast to small adjacent positive and negative peaks for trivial insulators. For large AFM order, it may be useful as a pure magneto-optical rotator due to sizable Kerr and Faraday angles and vanishing ellipticity. In the latter case, AFMTI shows much larger Kerr and Faraday angles and ellipticity than trivial insulators. Particularly, Kerr and Faraday angles of AFMTI are very sensitive to the change of spin configurations of ground state. These phenomena provides new features to distinguish AFMTI and trivial phases, and optially characterize the layered AFM materials.

This paper is organized as follows. In Sec.~\ref{sec:model}, we introduce model and methods, including model Hamiltonian and symmetries, Kerr and Faraday effects. The results for bilayer AFM insulators MnBi$_2$Te$_4$ are given in Sec.~\ref{sec:results}, including symmetry breaking by perpendicular electric field and external magnetic moment. Finally, conclusions and discussions are made in Sec.~\ref{sec:conclusion}.

\section{\label{sec:model}Model and methods}

\begin{figure}[t]
\centering \includegraphics[width=0.50\textwidth]{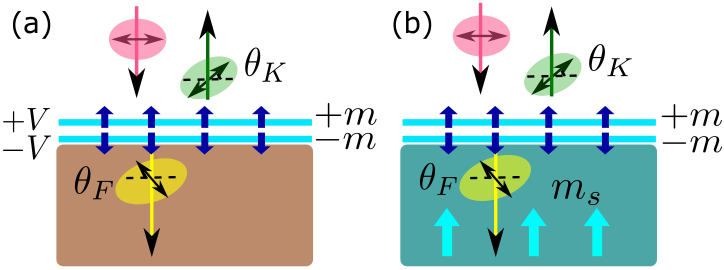}
\caption{Schematic illustration of magneto-optical Kerr and Faraday effects in bilayer AFM insulator MnBi$_2$Te$_4$ (a) with asymmetric potential $V$ induced by perpendicular electric field and (b) with external magnetic moment $m_s$. $\theta_{K}$ and $\theta_{F}$ refer to the Kerr and Faraday angles.}
\label{fig:setup}
\end{figure}

We consider a bilayer A-type AFM insulator MnBi$_2$Te$_4$ (layer number $n_z=2$) with out-of-plane magnetization. The tight-binding model is given by 
\begin{equation}
\begin{split}\label{model0}
\mathcal{H}&=\sum_{\bm k,z}c^+_{\bm k,z}[h_{a}(\bm k,z)+V(z)+m_s(z)]c_{\bm k,z}\\
&+[\sum_{\bm k,z}c^+_{\bm k,z}h_{b}(\bm k,z)c_{\bm k,z+1}+\mathrm{H.c.}],
\end{split}
\end{equation}
where the layer index $z=1,2$. The intralayer (interlayer) Hamiltonian $h_{a/b}(\bm k,z)$ read~\cite{zhang2019,zhangr2020,chen2021,ding2020,li2021,chen2022}
\begin{equation}
\begin{split}\label{model1}
h_{a}(\bm k,z) &=   M(\bm k)\sigma_z + A_2\sum_{\alpha=x,y}\sin k_{\alpha}s_{\alpha}\sigma_x
+m(z)s_z,\\
h_{b}(\bm k,z) &=B_1\sigma_z-i\frac{A_1}{2}s_z\sigma_x.
\end{split}
\end{equation}
Pauli matrices $s$ and $\sigma$ act in the spin and orbital subspaces, respectively. $M(\bm k)=M_0-2B_1-2B_2(2-\cos k_x-\cos k_y)$. $M_0$, $B_i$ and $A_i$ are model parameters, where $i=1,2$. $m(z)=m(-1)^z$ is the AFM order in each layer. $V(z)=V[2z-n_z-1]$ and $m_s(z)=m_ss_z$ are the layer-dependent potential and layer-independent magnetism due to external electric field and magnetic moment of substrates, respectively (see Fig.~\ref{fig:setup}). When $V=m_s=0$, the system describes a trivial insulator for $M_0<0$, and an AFMTI for $M_0>0$~\cite{chen2021,chen2022}.

The space group of MnBi$_2$Te$_4$ is $D^5_{3d}$ (No.166) in the absence of magnetic order. The presence of AFM order $m(z)$ breaks both time-reversal symmetry $\mathcal{T}=is_y\mathcal{K}$ and inversion symmetry $\mathcal{P}=\sigma_z$, whereas it preserves the combined $\mathcal{P}\mathcal{T}$ symmetry (see Table \ref{tab:sym}). Since $(\mathcal{P}\mathcal{T})^2=-1$, there is Kramer's degeneracy at each momentum of Brillouin zone. In addition, the system preserves the combined symmetries $\mathcal{M}_{1}=\mathcal{M}_{x}\mathcal{M}_{z}$ and $\mathcal{M}_{2}=\mathcal{M}_{y}\mathcal{M}_{z}$, where $\mathcal{M}_{\alpha=x,y,z}$ is the mirror reflection symmetry operation $\alpha\rightarrow-\alpha$. These combined mirror symmetries correspond to the combined rotational symmetries $\mathcal{C}_{2x}\mathcal{C}_{2z}$ and $\mathcal{C}_{2y}\mathcal{C}_{2z}$ in the system~\cite{zhang2019} ($\mathcal{C}_{2\alpha}$ is the twofold rotational symmetry operation about $\alpha$ axis, $\alpha=x,y,z$). When either $V(z)$ or $m_s(z)$ term is introduced, both $\mathcal{P}\mathcal{T}$ and $\mathcal{M}_{1/2}$ symmetries are broken.

\renewcommand\arraystretch{1.5}

\begin{table}[b]
\caption{ Symmetries of even-layered AFM Hamiltonian $h_{a/b}(\bm k,z)$ satisfying $\hat{O}^{-1}h_{a/b}(\bm k,z)\hat{O}=h(\bm k',z')$, where $\hat{O}$ is the symmetry operator. $\surd$ ($\times$) implies the presence (absence) of the symmetry relations. $(\bm k',z')=\hat{O}^{-1}(\bm k,z)$ is indicated and $\bar{z}=n_z+1-z$.}
\label{tab:sym}%
\begin{ruledtabular}
\begin{tabular}{*6c} 
 & $(\bm k',z')$  & $h_a(\bm k,z)$ & $h_b(\bm k,z)$ & $V(z)$  &  $m_s(z)$\\
  \hline
$\mathcal{T}=is_y\mathcal{K}$ & $(-\bm k,z)$  & $\times$ & $\surd$  &  $\surd$  &  $\times$ \\
$\mathcal{P}=\sigma_z$ & $(-\bm k,\bar{z})$  & $\times$ & $\surd$  &  $\times$  &  $\surd$ \\
$\mathcal{P}\mathcal{T}=is_y\sigma_z\mathcal{K}$ & $(\bm k,\bar{z})$  & $\surd$ & $\surd$ & $\times$  &  $\times$ \\
$\mathcal{M}_1=s_x\sigma_z$ & $(-k_x,k_y,\bar{z})$  & $\surd$ & $\surd$  & $\times$  &  $\times$ \\
$\mathcal{M}_2=s_y\sigma_z$ & $(k_x,-k_y,\bar{z})$  & $\surd$ & $\surd$  & $\times$  &  $\times$ \\
\end{tabular}
\end{ruledtabular}
\end{table}

When a linearly polarized light is incident into magnetic materials, there are phase differences between right-circularly and left-circularly polarized components for reflected and transmitted light. These lead to rotations of polarization axis (see Fig.~\ref{fig:setup}), namely magneto-optical Kerr and Faraday effect, respectively.  Such phenomena are not restricted to bulk samples, but widely used in studying spin structures of few-layer materials~\cite{nandkishore2011,crassee2011,shimano2013,sivadas2016,huang2017,gong2017}. Since few-layer materials are much thinner than the light wavelength, it is convenient to treat them as interfaces between vacuum and substrate. As a result, Kerr and Faraday angles are given by~\cite{freiser1968,tse2010,gorbar2012,tse2011,catarina2020}
\begin{equation}
\begin{split} 
\theta_K &= \frac{1}{2}[\arg(r_-)-\arg(r_+)],\\
\theta_F &= \frac{1}{2}[\arg(t_-)-\arg(t_+)],
\end{split}
\end{equation}
where the reflection and transmission coefficients for the circularly-polarized light read ($\pm$ for left- and right-hand)
\begin{equation}
\begin{split} 
r_{\pm} &=  \frac{1-\sqrt{\frac{\epsilon_r}{\mu_r}}-Z_0\sigma_{\pm}}
{1+\sqrt{\frac{\epsilon_r}{\mu_r}}+Z_0\sigma_{\pm}},\   \ 
t_{\pm} = \frac{2}
{1+\sqrt{\frac{\epsilon_r}{\mu_r}}+Z_0\sigma_{\pm}}.
\end{split}
\end{equation}
$Z_0=\sqrt{\frac{\mu_0}{\epsilon_0}}\approx376.7\Omega$ is the impedance of vacuum, where $\epsilon_0$ and $\mu_0$ are vacuum permittivity and permeability, respectively. The magnetic permeability $\mu_r=1$ and dielectric constant $\epsilon_r=4$ for SiO$_2$ substrate~\cite{tse2011,catarina2020}. Here $\sigma_{\pm}=$ $\sigma_{xx}$ $\pm$ $i\sigma_{xy}$, where $\sigma_{xx}$ ($\sigma_{xy}$) is the longitudinal (Hall) optical conductivity tensor. Additionally, the Kerr and Faraday ellipticity $\gamma_K$, $\gamma_F$ can be defined~\cite{freiser1968,catarina2020}
\begin{equation}
\begin{split}
\tan \gamma_K=\frac{|r_-|-|r_+|}{|r_-|+|r_+|},\   \  \tan \gamma_F=\frac{|t_-|-|t_+|}{|t_-|+|t_+|}.
\end{split}
\end{equation}

By combining $\theta_K$, $\theta_F$ and $\gamma_K$, $\gamma_F$, complex Kerr and Faraday angles can be introduced~\cite{kahn1969}
\begin{equation}
\begin{split}
\phi_K &= \theta_K+i\gamma_K,\   \  \phi_F = \theta_F+i\gamma_F.
\end{split}
\end{equation}

By Kubo formula, the finite-frequency conductivity tensor follows~\cite{mahan2000}
\begin{equation}
\begin{split}\label{conductivity}
&\sigma_{\alpha \beta}(\omega)=\\
&i\hbar\sum_{\mu \mu'}\int\frac{d^2\bm k}{(2\pi)^2} \frac{f_{\bm k\mu}-f_{\bm k \mu'}}{\epsilon_{\bm k \mu}-\epsilon _{\bm k\mu'}} \frac{\langle \bm k, \mu|j_\alpha| \bm k, \mu' \rangle \langle \bm k, \mu'|j_\beta| \bm k, \mu \rangle}{\omega+\epsilon_{\bm k \mu}-\epsilon _{\bm k\mu'}+i\hbar/2\tau_s},
\end{split}
\end{equation}
where $\epsilon_{\bm k\mu}$ and $|\bm k,\mu\rangle$ are the eigenvalues and eigenstates of Hamiltonian $\mathcal{H}$. $\bm k=(k_x,k_y)$ is the wave vector and $\mu, \mu'=\{1,2,\cdots,4n_z\}$ are band indices. At zero temperature, the Fermi-Dirac distribution $f_{\bm k \mu}=1/[1+\exp((\epsilon_{\bm k\mu}-\epsilon_F)/k_BT)]=\Theta(\epsilon_F-\epsilon_{\bm k\mu})$, where $\epsilon_F$ is the Fermi energy and $\Theta(...)$ is the Heaviside function. $\omega$ is the photon energy and $\tau_s$ is the relaxation time. The current operator $j_{\alpha}=(e/\hbar)\partial h(\bm k)/\partial k_{\alpha}$, $\alpha=x,y$, where the effective Hamiltonian is established by $h(\bm k)=\bigoplus_{z=1}^{n_z=2}[h_a(\bm k,z)+h_b(\bm k,z)+V(z)+m_s(z)]$ from Eq. (\ref{model1}). 

Note that for even-layered MnBi$_2$Te$_4$ with $V=m_s=0$, the Hall conductivity $\sigma_{xy}(\omega)=0$ due to the presence of combined mirror symmetries $\mathcal{M}_{1/2}$ rather than $\mathcal{P}\mathcal{T}$ symmetry, as shown in Appendix \ref{sec:sym}. When $V\neq0$ or $m_s\neq0$, neither of symmetry $\mathcal{M}_{1}$ and $\mathcal{M}_{2}$ is preserved, giving rise to nonzero $\sigma_{xy}$, and hence the Kerr and Faraday effect.

\begin{figure}[t]
\centering \includegraphics[width=0.49\textwidth]{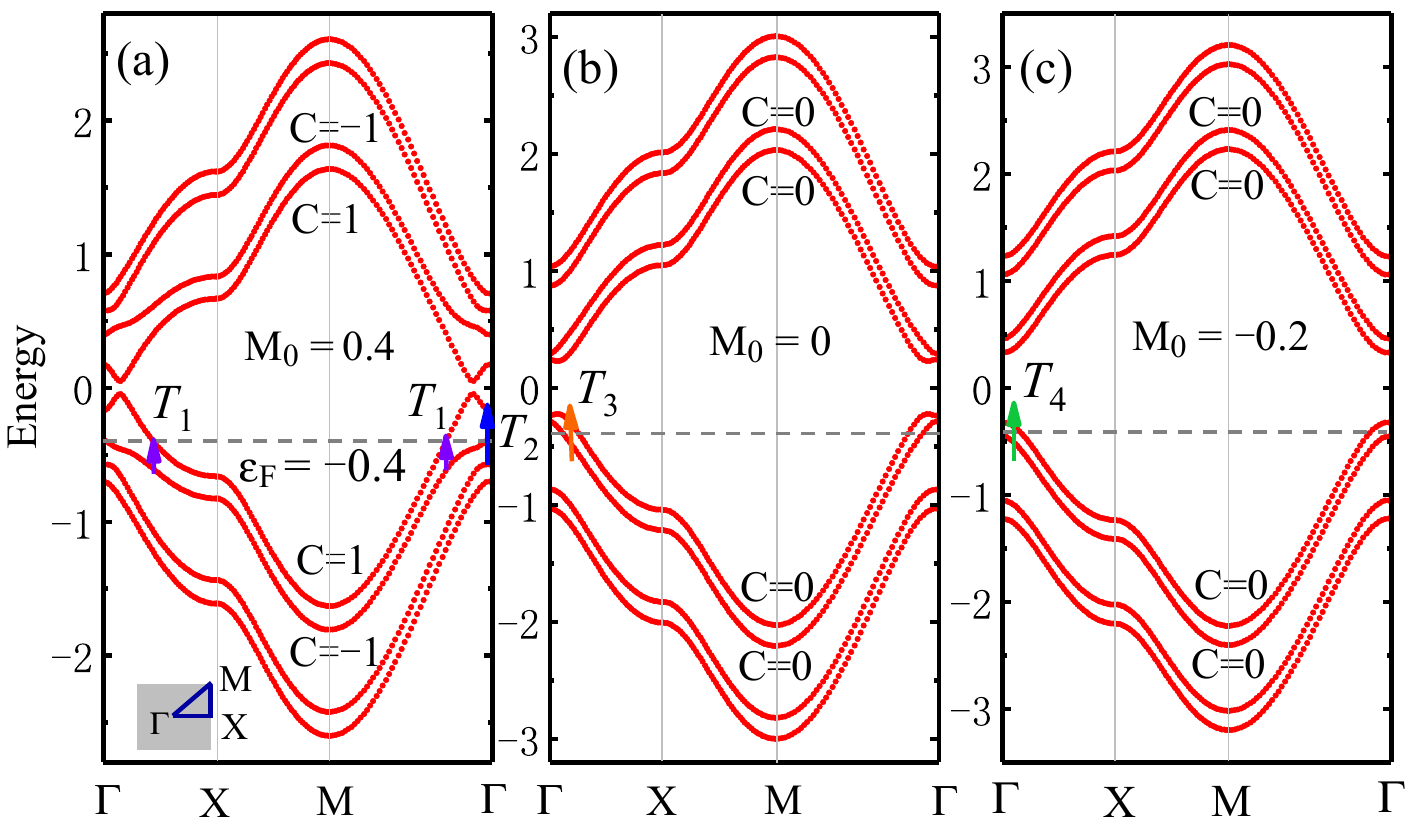}
\caption{ Band dispersions of bilayer AFM insulator MnBi$_2$Te$_4$ with gate voltage $V$ for parameters (a) $M_0=0.4$, (b) $M_0=0$ and (c) $M_0=-0.2$. The dispersions are plotted along the high-symmetry lines of Brillouin zone, as indicated in the inset of (a). In the absence of potential $V$, $M_0=0.4$ and $-0.2$ correspond to AFMTI and trivial insulator, respectively. In the presence of potential $V$, the Chern number of each band is labeled. The optical interband transition threshold $T_{i=1,2,3,4}$ for low photon energy are depicted by arrows.  Parameters: $A_1=A_2=0.55$, $B_1=B_2=0.25$~\cite{li2021,chen2022}, $m=0.12$, $V=0.3$, $\epsilon_F=-0.4$.}
\label{fig:layerAFM1}
\end{figure}

\section{\label{sec:results}Results}

\subsection{Symmetry breaking by perpendicular electric field}

In this section, we show numerical results of optical conductivities, Kerr and Faraday angles and ellipticity for bilayer AFM insulator due to symmetry breaking by perpendicular electric field. Band dispersions are plotted in Fig.~\ref{fig:layerAFM1} for parameters $M_0=0.4$, $0$ and $-0.2$, respectively. In the absence of $V$, $M_0=0.4$ and $-0.2$ correspond to AFMTI and trivial insulator, respectively. In the presence of $V$, the band edge of AFMTI ($M_0=0.4$) is moved away from the $\Gamma$ point, whereas trivial insulator ($M_0=-0.2$) shows monotonic dispersion around the $\Gamma$ point. The non-monotonic dispersion of AFMTI arises from the band inversion of two highest bands below (or two lowest bands above) zero energy. This is evidenced by the value of Chern number of each band (see Fig.~\ref{fig:layerAFM1} (a)), from which one can understand the origin of band inversion. Numerical results suggest that the topological phase transition takes place for $M_0=0.07$, when these bands touch at the $\Gamma$ point. For clarity, we plot the dispersion for $M_0=0$ in Fig.~\ref{fig:layerAFM1} (b), whose dispersion is quite similar to $M_0=0.07$. When the Fermi energy $\epsilon_F=0$, the summation of Chern number $C$ of occupied bands vanishes, leading to zero static Hall conductivity $\sigma_{xy}(\omega=0)$. To avoid this, we set the Fermi level $\epsilon_F=-0.4$. In this case, we label the optical interband transition threshold $T_{i=1,2,3,4}$ by arrows in Fig.~\ref{fig:layerAFM1}, which are useful in understanding the behaviors of optical conductivities, Kerr and Faraday angles at low photon energy $\omega$.

\begin{figure}[t]
\centering \includegraphics[width=0.48\textwidth]{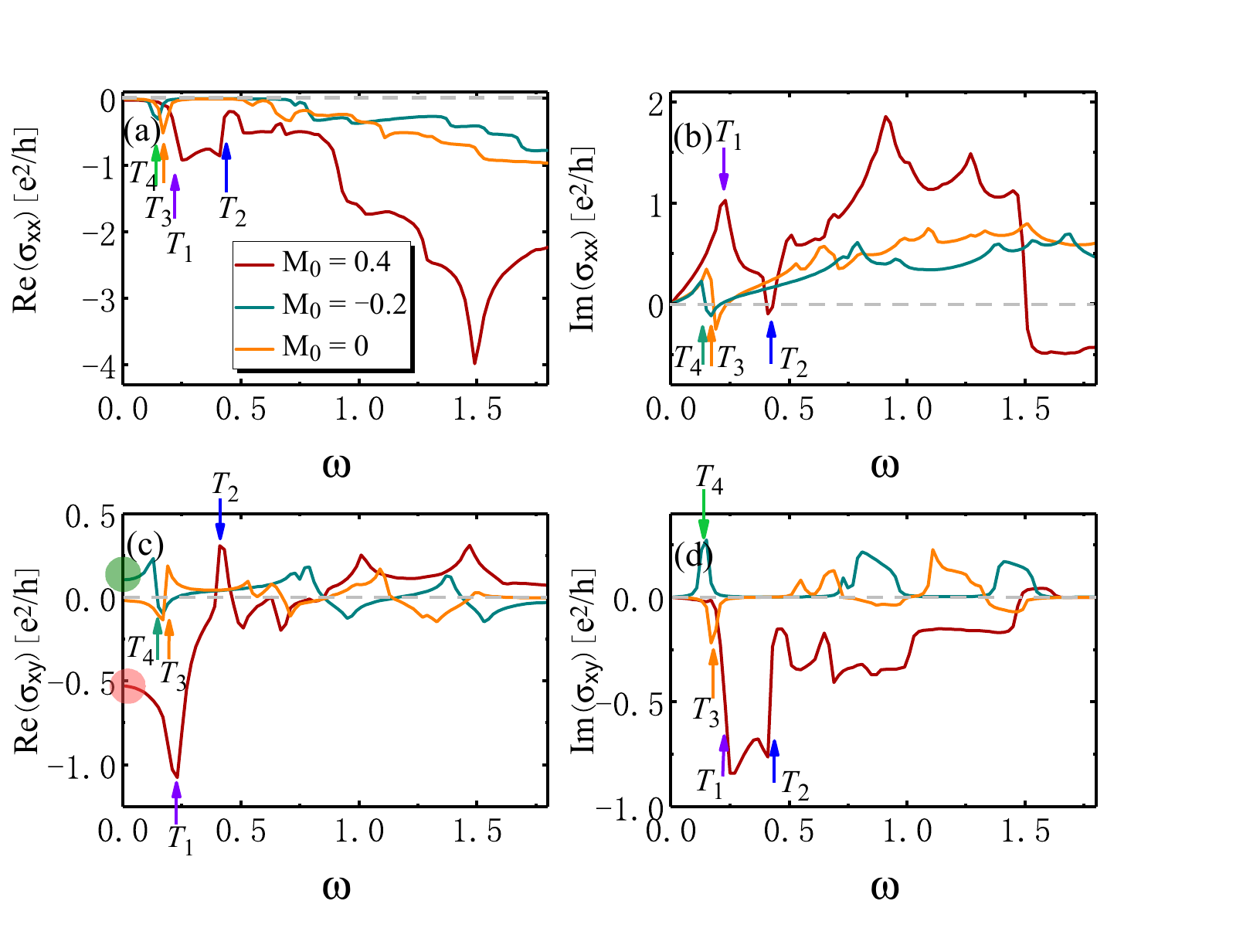}
\caption{ Real and imaginary parts of optical conductivities (a)-(b) $\sigma_{xx}$ and (c)-(d) $\sigma_{xy}$ (in units of $e^2/h$) as functions of photon energy $\omega$ for bilayer AFM insulator MnBi$_2$Te$_4$. The arrows refer to the optical interband transitions $T_{i=1,2,3,4}$ as shown in Fig.~\ref{fig:layerAFM1}. The values of $\mathrm{Re}(\sigma_{xy})$ in the low-energy limit are highlighted in red and green region in (c). Parameters: $m=0.12$, $V=0.3$, $\epsilon_F=-0.4$, $\epsilon_r=4$~\cite{tse2011,catarina2020}, $\mu_r=1$, $\hbar/\tau_s=0.02$.}
\label{fig:layerAFM2}
\end{figure}

The real and imaginary parts of optical conductivities $\sigma_{xx}$ and $\sigma_{xy}$ are shown in Fig.~\ref{fig:layerAFM2}. Here we only focus on the features at low photon energy $\omega$. One prominent feature is that the magnitude of $\sigma_{xx}$ and $\sigma_{xy}$ for AFMTI ($M_0=0.4$) is much larger than trivial insulator ($M_0=-0.2$), as a result of band inversion. The behaviors between $\mathrm{Re}[\sigma_{xx}]$ and $\mathrm{Im}[\sigma_{xy}]$ (or $\mathrm{Im}[\sigma_{xx}]$ and $\mathrm{Re}[\sigma_{xy}]$) are similar, whereas the peaks (jumps) of $\mathrm{Re}[\sigma_{xx}]$ and $\mathrm{Im}[\sigma_{xy}]$ correspond to the jumps (peaks) of $\mathrm{Im}[\sigma_{xx}]$ and $\mathrm{Re}[\sigma_{xy}]$ due to the Kramers-Kr$\ddot{o}$nig relations~\cite{giuliani2005}. For AFMTI ($M_0=0.4$), $\mathrm{Re}[\sigma_{xy}]$ shows sharp peaks at the interband transition threshold $T_1$ and $T_2$, as shown in Fig.~\ref{fig:layerAFM2} (c). By contrast, for trivial insulator ($M_0=-0.2$), $\mathrm{Re}[\sigma_{xy}]$ exhibits two small adjacent positive and negative peaks, which arise from the appearance and disappearance of transition $T_{4}$. The peaks are close to each other since the energy window of the transition is narrow as a result of monotonic band dispersion (see Fig.~\ref{fig:layerAFM1} (c)). The value of $\mathrm{Re}(\sigma_{xy})$ in the static limit $\omega\rightarrow0$ is highlighted in Fig.~\ref{fig:layerAFM2} (c), which is proportional to the integral of Berry curvature of occupied bands.

\begin{figure}[t]
\centering \includegraphics[width=0.49\textwidth]{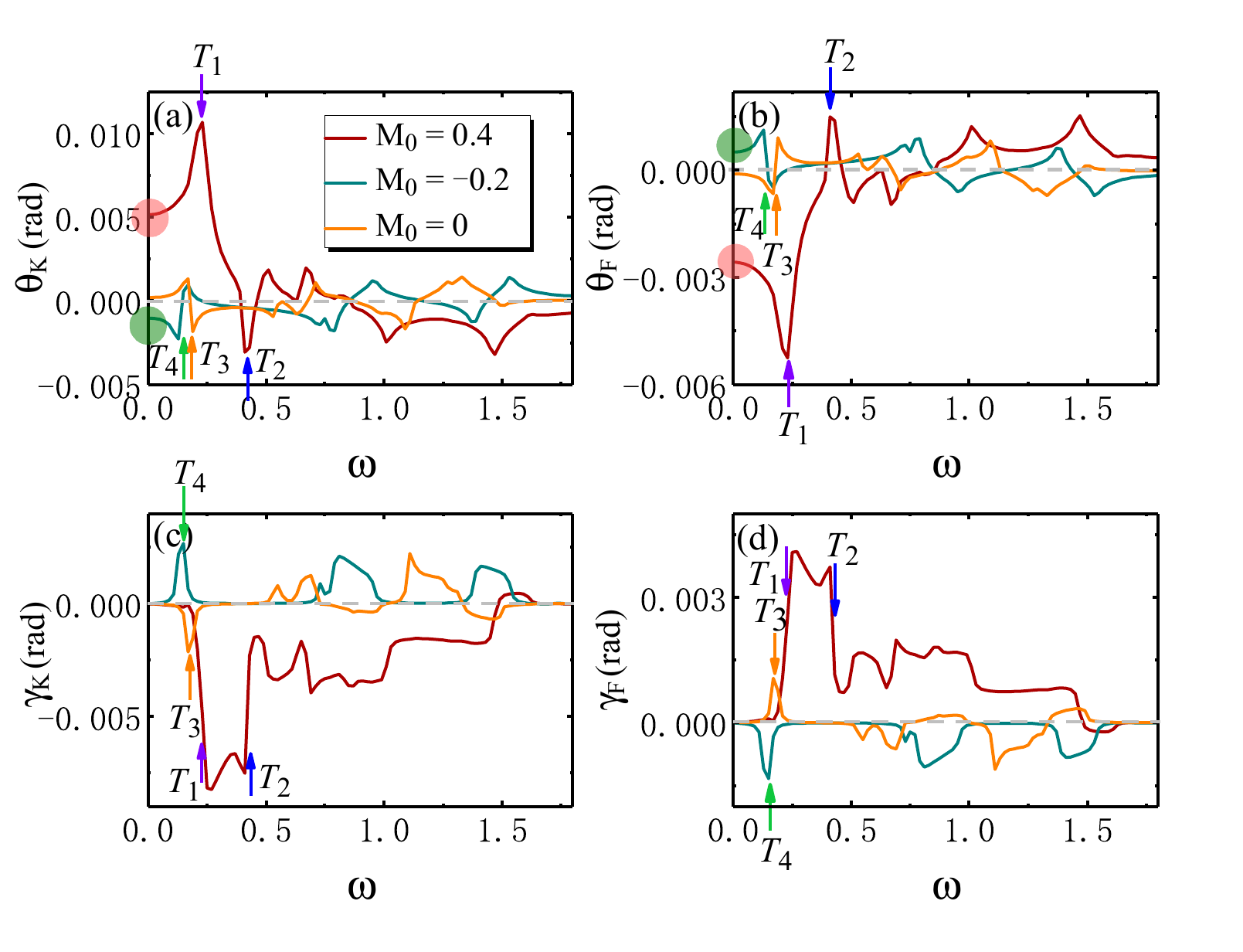}
\caption{ (a) Kerr and (c) Faraday angles and (b) Kerr and (d) Faraday ellipticity as functions of photon energy $\omega$ for bilayer AFM insulator MnBi$_2$Te$_4$. The arrows refer to the optical interband transitions $T_{i=1,2,3,4}$ as shown in Fig.~\ref{fig:layerAFM1}. The values of $\theta_K$ and $\theta_F$ in the low-energy limit are highlighted in (a) and (b). Parameters: $m=0.12$, $V=0.3$, $\epsilon_F=-0.4$, $\epsilon_r=4$~\cite{tse2011,catarina2020}, $\mu_r=1$, $\hbar/\tau_s=0.02$.}
\label{fig:layerAFM3}
\end{figure}

The Kerr and Faraday angles and ellipticity are plotted in Fig.~\ref{fig:layerAFM3}. $\theta_K$ and $\theta_F$ (also $\gamma_K$ and $\gamma_F$) are complementary to each other since they refer to the reflected and transmitted light, respectively. We find that the magnitude of $\theta_{K/F}$ and $\gamma_{K/F}$ is largely enhanced for AFMTI ($M_0=0.4$), as compared to trivial insulator ($M_0=-0.2$). Moreover, it is clear that $\theta_{K/F}$ inherits the main features from $\mathrm{Re}(\sigma_{xy})$, and shows sharp peaks at the interband absorption threshold $T_1$ and $T_2$ for $M_0=0.4$. On the other hand, $\gamma_{K/F}$ inherits the main features from $\mathrm{Im}(\sigma_{xy})$, and shows abrupt jumps at transitions $T_1$ and $T_2$ for $M_0=0.4$. Therefore the measurement of Kerr and Faraday angles, ellipticity, and also optical conductivities, provides a way to distinguish between AFMTI and trivial insulator.

\begin{figure}[htbp]
\centering \includegraphics[width=0.49\textwidth]{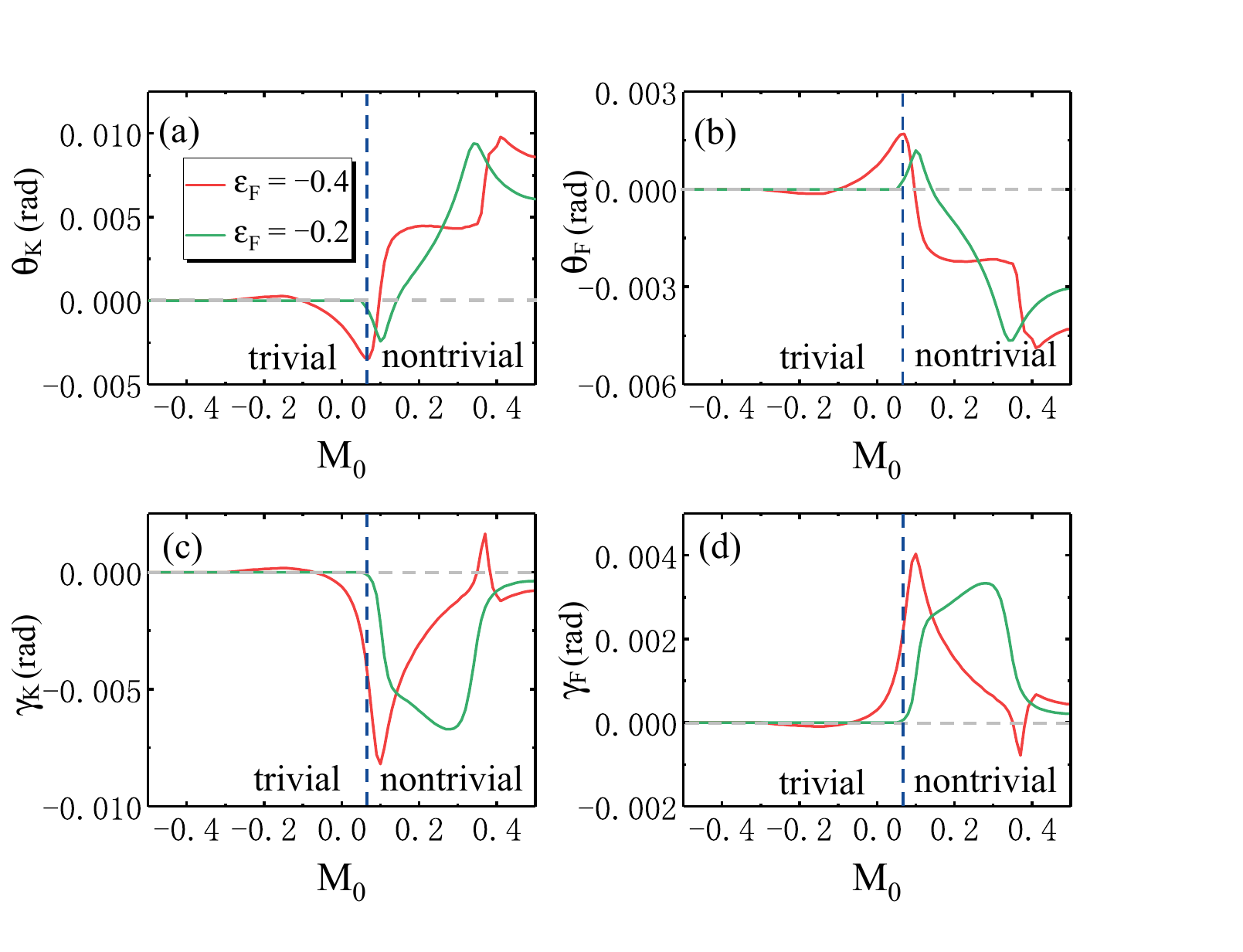}
\caption{ (a) Kerr and (c) Faraday angles and (b) Kerr and (d) Faraday ellipticity as functions of $M_0$ for bilayer AFM insulator MnBi$_2$Te$_4$. Topological phase transition occurs for $M_0=0.07$, as indicated by the blue dashed line. Red and green lines correspond to Fermi energy $\epsilon_F=-0.4$ and $-0.2$, respectively. Parameters: $m=0.12$, $V=0.3$, $\omega=0.2$, $\epsilon_r=4$~\cite{tse2011,catarina2020}, $\mu_r=1$, $\hbar/\tau_s=0.02$.}
\label{fig:layerAFM4}
\end{figure}

Since tuning $M_0$ can drive a topological phase transition between AFMTI and trivial insulator, we show $M_0$ dependence of Kerr and Faraday angles and ellipticity in Fig.~\ref{fig:layerAFM4}. It is clear that nontrivial AFMTI has much larger magnitude of $\theta_{K/F}$ and $\gamma_{K/F}$ than trivial phase. The topological phase transition occurs when $M_0=0.07$, where $\theta_{K/F}$ reaches a peak for Fermi energy $\epsilon_F=-0.4$ while $\gamma_{K/F}$ is slightly away from the peak. For higher Fermi level $\epsilon_F=-0.2$, the peaks are further moved away from the transition point since larger $M_0$ is required to reduce the band gap and enhance the valence band edge.

\begin{figure}[htbp]
\centering \includegraphics[width=0.49\textwidth]{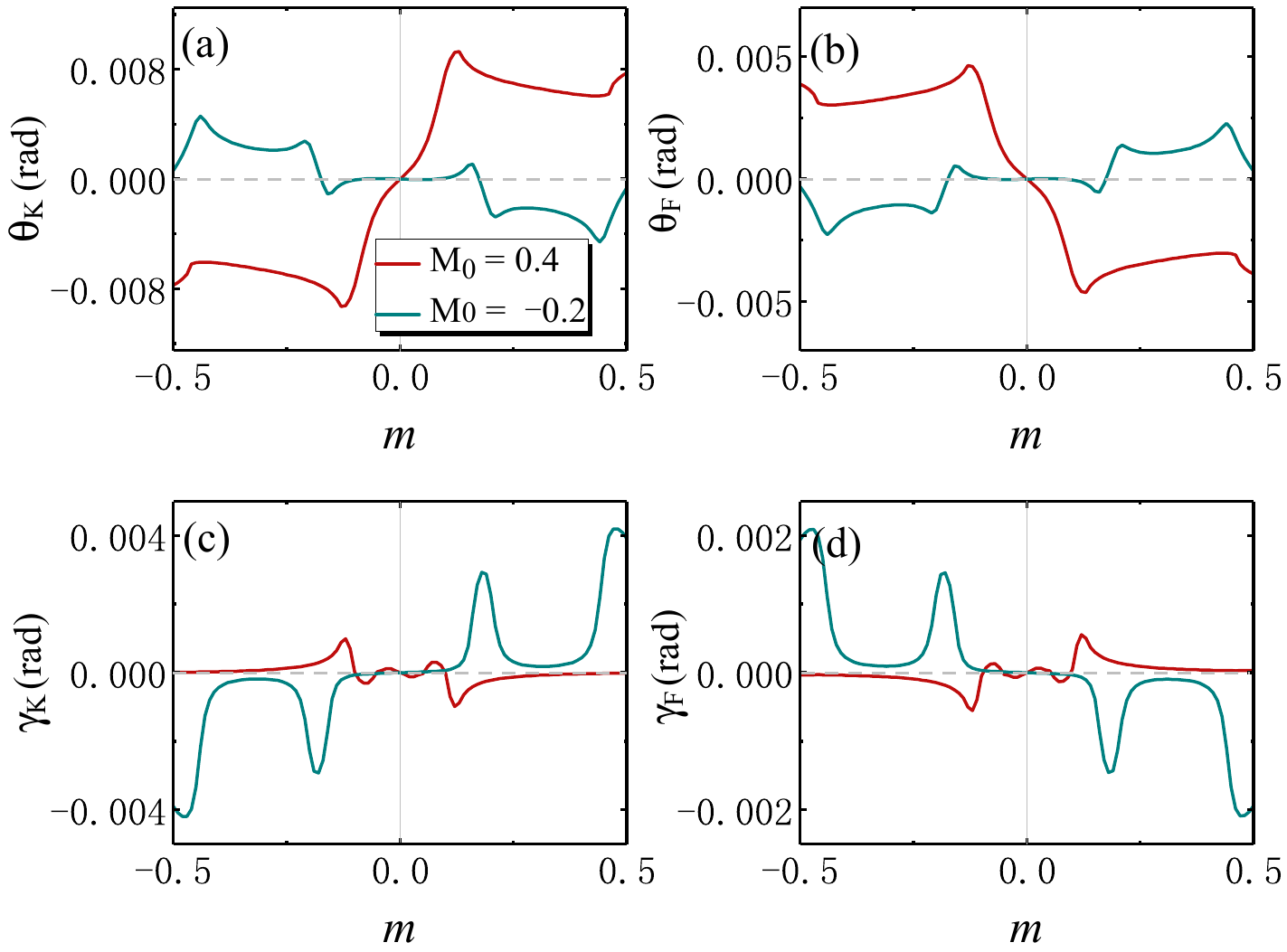}
\caption{ (a) Kerr and (c) Faraday angles and (b) Kerr and (d) Faraday ellipticity as functions of AFM order $m$ for bilayer AFM insulator MnBi$_2$Te$_4$. Red and green lines correspond to AFMTI ($M_0=0.4$) and trivial insulator ($M_0=-0.2$), respectively. Parameters: $V=0.3$, $\omega=0.2$, $\epsilon_F=-0.4$, $\epsilon_r=4$~\cite{tse2011,catarina2020}, $\mu_r=1$, $\hbar/\tau_s=0.02$.}
\label{fig:layerAFM5}
\end{figure}

AFM order $m$ dependence of Kerr and Faraday angles and ellipticity is plotted in Fig.~\ref{fig:layerAFM5}, where $\theta_{K/F}$ and $\gamma_{K/F}$ flip signs under the inversion of magnetic moment $m$. For AFMTI ($M_0=0.4$), the magnitude of $\theta_{K/F}$ increases when $m$ increases, then it almost saturates when only one valence band is intersected by the Fermi level. By contrast, the magnitude of $\theta_{K/F}$ for trivial insulator ($M_0=-0.2$) is much smaller. Interestingly, for AFMTI, when the magnetic moment $m$ is large, $\theta_{K/F}$ has sizable value with almost vanishing ellipticity. This means the system can be used as a pure magneto-optical rotator.

\begin{figure}[t]
\centering \includegraphics[width=0.49\textwidth]{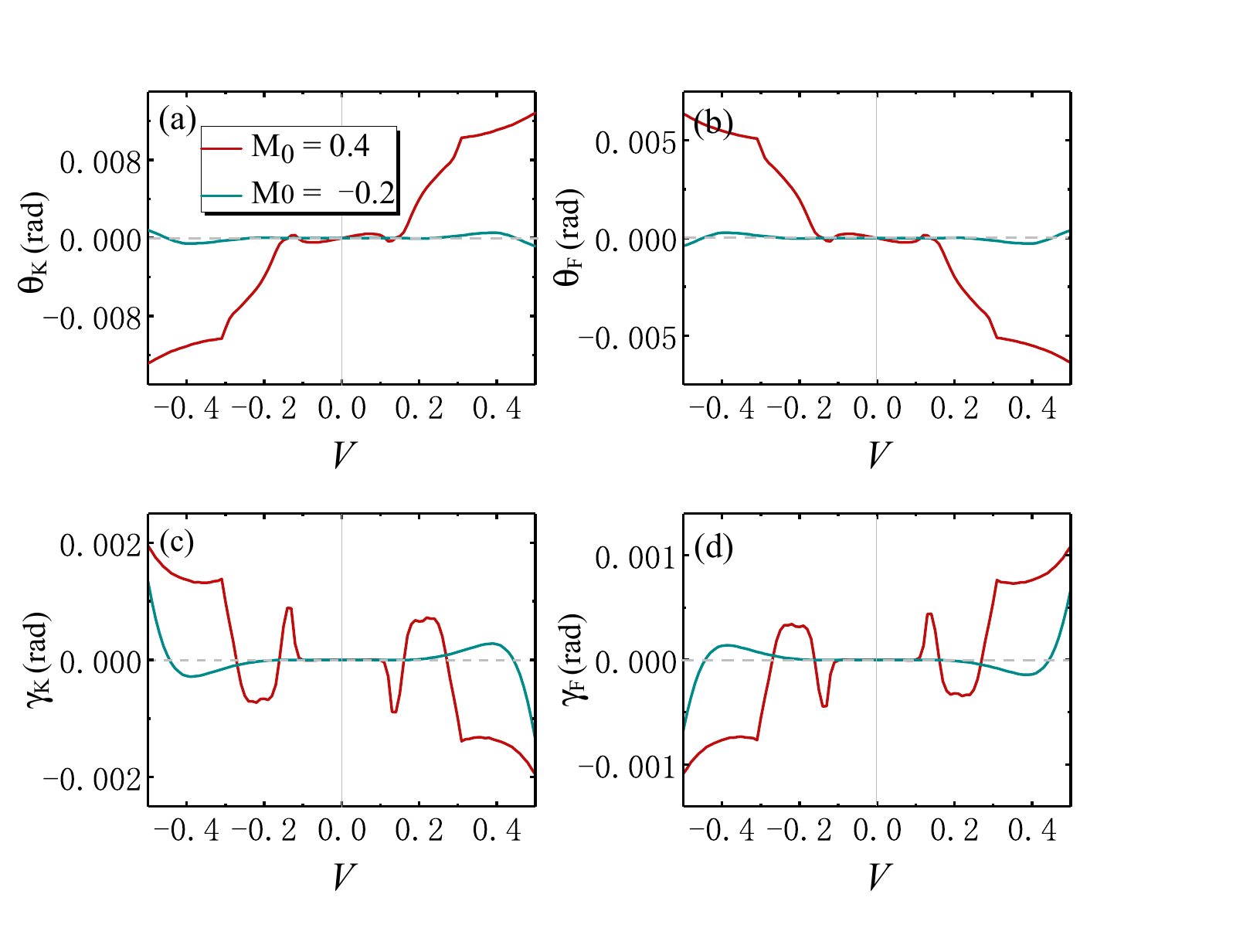}
\caption{ (a) Kerr and (c) Faraday angles and (b) Kerr and (d) Faraday ellipticity as functions of gate voltage $V$ for bilayer AFM insulator MnBi$_2$Te$_4$. Red and green lines correspond to AFMTI ($M_0=0.4$) and trivial insulator ($M_0=-0.2$), respectively. Parameters: $m=0.12$, $\omega=0.2$, $\epsilon_F=-0.4$, $\epsilon_r=4$~\cite{tse2011,catarina2020}, $\mu_r=1$, $\hbar/\tau_s=0.02$.}
\label{fig:layerAFM6}
\end{figure}

Tuning gate voltage $V$ is more feasible in experiments, thus we provide the $V$ dependence of Kerr and Faraday angles and ellipticity in Fig.~\ref{fig:layerAFM6}. Similar to the reciprocal behaviors under magnetic moment, both $\theta_{K/F}$ and $\gamma_{K/F}$ flip signs under the inversion of gate voltage $V$. For AFMTI ($M_0=0.4$), we find that $\theta_{K/F}$ is negligibly small when $|V|<0.12$, and it is largely enhanced when $|V|>0.12$. By contrast, for trivial insulator ($M_0=-0.2$), $\theta_{K/F}$ is always tiny. The turning point $V=\pm0.12$ for AFMTI is due to the occurrence of non-monotonic band dispersions near the $\Gamma$ point, as shown in Fig.~\ref{fig:layerAFM1} (a).

\begin{figure}[t]
\centering \includegraphics[width=0.49\textwidth]{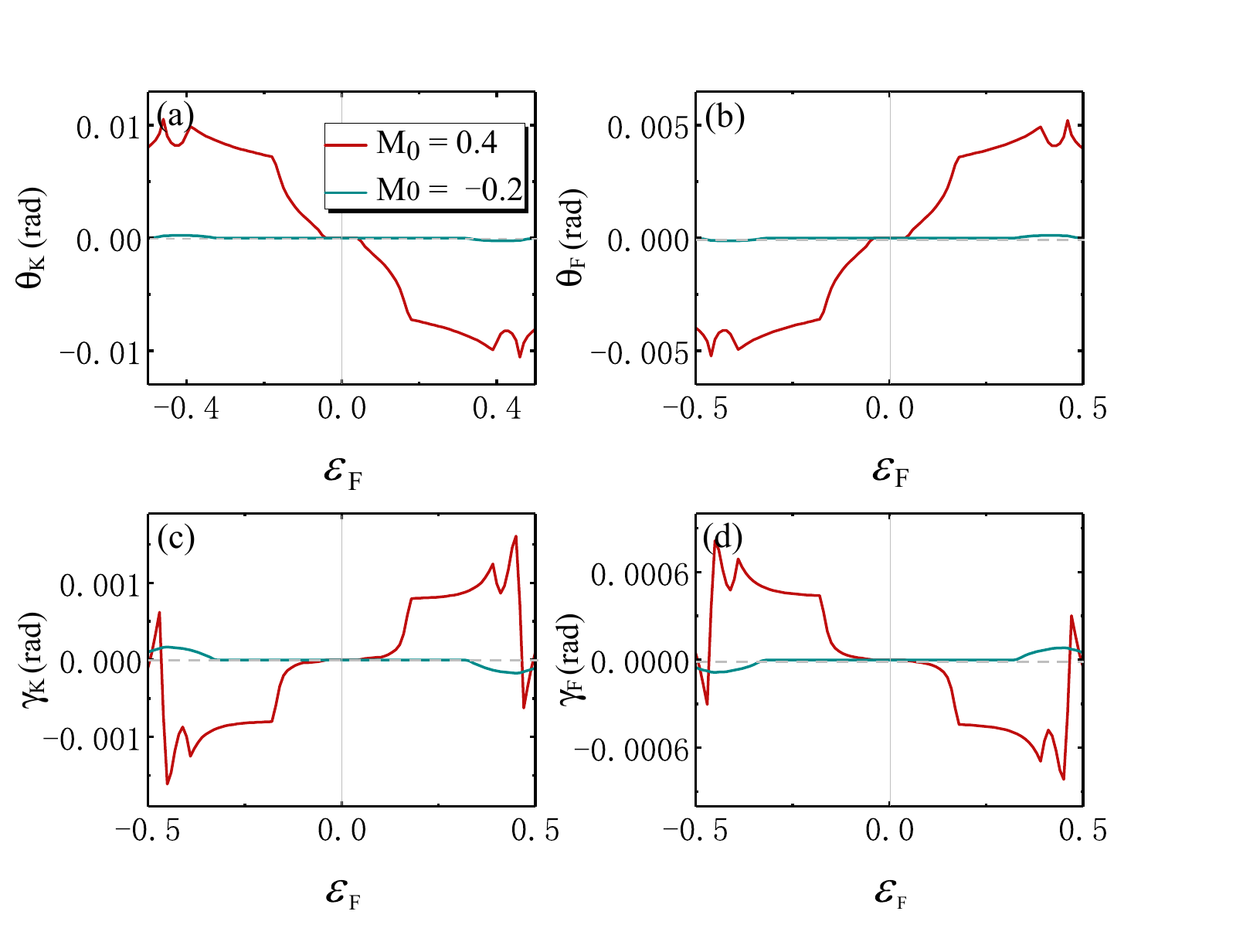}
\caption{ (a) Kerr and (c) Faraday angles and (b) Kerr and (d) Faraday ellipticity as functions of Fermi energy $\epsilon_F$ for bilayer AFM insulator MnBi$_2$Te$_4$. Red and green lines correspond to AFMTI ($M_0=0.4$) and trivial insulator ($M_0=-0.2$), respectively. Parameters: $m=0.12$, $V=0.3$, $\omega=0.2$, $\epsilon_r=4$~\cite{tse2011,catarina2020}, $\mu_r=1$, $\hbar/\tau_s=0.02$.}
\label{fig:layerAFM7}
\end{figure}

Another tunable parameter in experiments is the Fermi energy $\epsilon_F$, whose dependence of Kerr and Faraday angles and ellipticity is shown in Fig.~\ref{fig:layerAFM7}. Both $\theta_{K/F}$ and $\gamma_{K/F}$ flip signs under the inversion of Fermi energy $\epsilon_F$. When $|\epsilon_F|>0.05$, the Fermi level starts to intersect the bands, giving rise to largely enhanced $\theta_{K/F}$ and $\gamma_{K/F}$ for AFMTI while vanishingly small value for trivial insulator. This provides a new feature to distinguish between AFMTI and trivial insulator.

\subsection{Symmetry breaking by external magnetic moment $m_s$}

\begin{figure}[b]
\centering \includegraphics[width=0.46\textwidth]{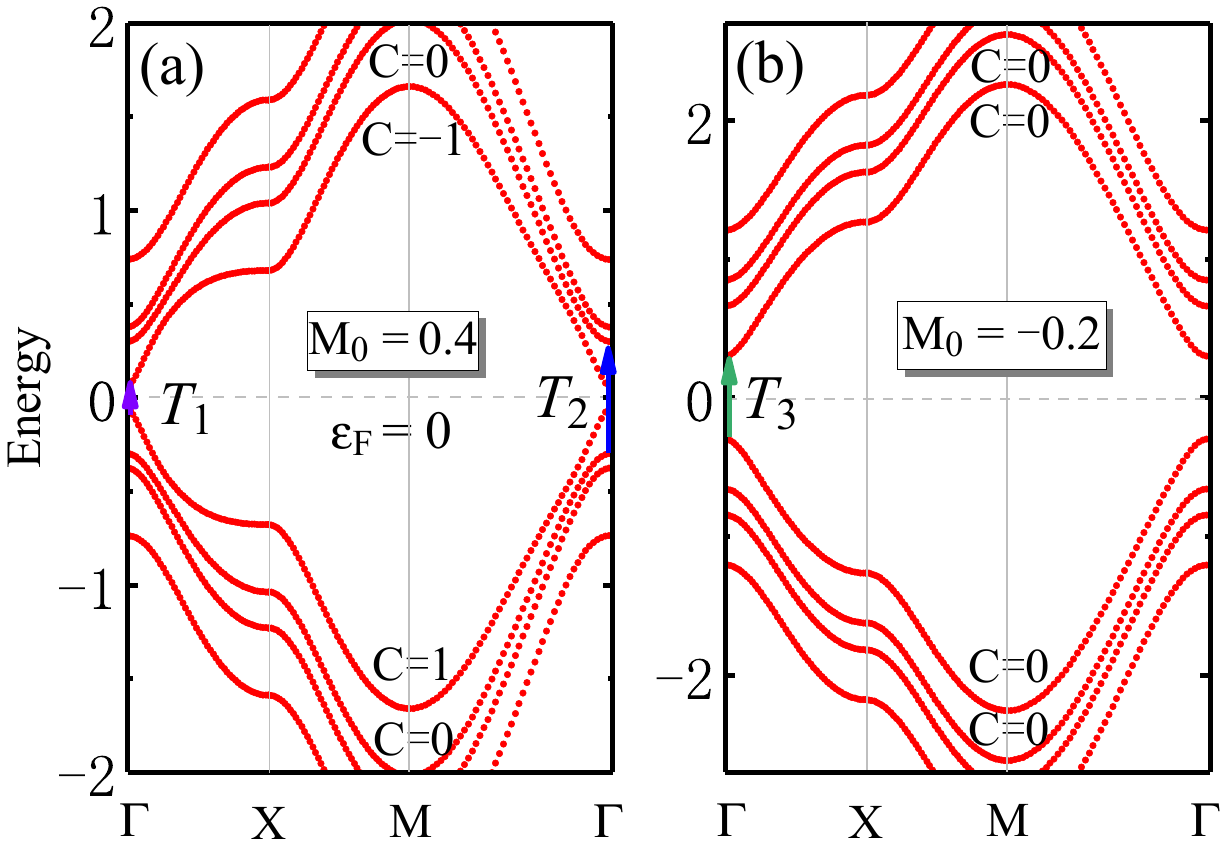}
\caption{ Band dispersions of bilayer AFM insulator MnBi$_2$Te$_4$ with external magnetic moment $m_s$ for parameters (a) $M_0=0.4$ and (b) $M_0=-0.2$. In the absence of magnetic moment $m_s$, $M_0=0.4$ and $-0.2$ correspond to AFMTI and trivial insulator, respectively. In the presence of $m_s$, the Chern number of each band is labeled. The optical interband transition threshold $T_{i=1,2,3}$ for low-energy photons are depicted by arrows.  Parameters: $A_1=A_2=0.55$, $B_1=B_2=0.25$~\cite{li2021,chen2022}, $m=0.12$, $m_s=0.18$, $\epsilon_F=0$.}
\label{fig:layerFIM1}
\end{figure}

In this section, we show numerical results of Kerr and Faraday angles and ellipticity due to symmetry breaking by external magnetic moment $m_s$. Band dispersions are plotted in Fig.~\ref{fig:layerFIM1} for parameters $M_0=0.4$ and $-0.2$, respectively. In the presence of $m_s$, AFMTI ($M_0=0.4$) become Chern insulators, as indicated by the Chern number of each band. Here, for simplicity, we set the Fermi energy $\epsilon_F=0$. Some crucial interband transition threshold $T_{i=1,2,3}$ are labeled by arrows.

\begin{figure}[t]
\centering \includegraphics[width=0.49\textwidth]{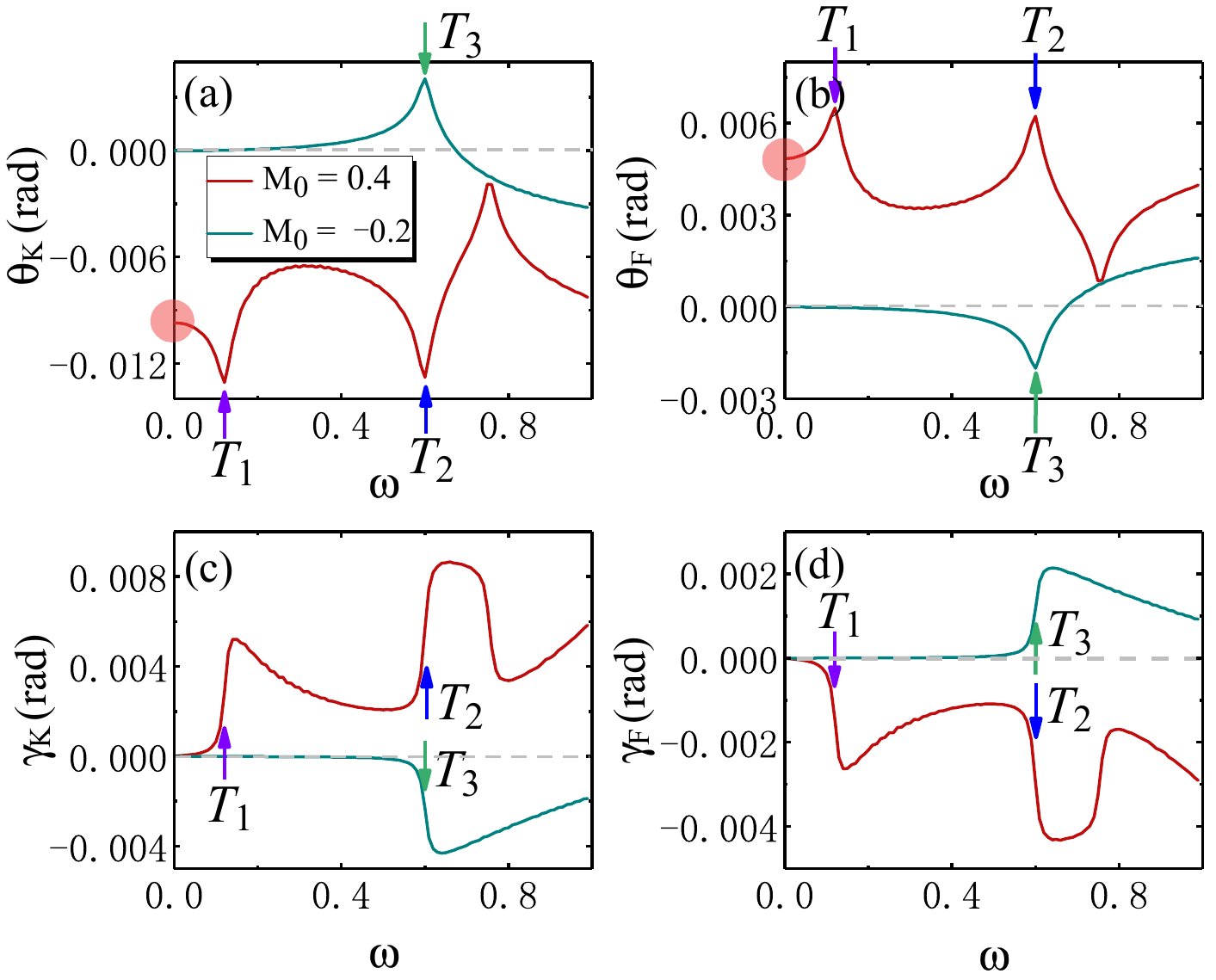}
\caption{ (a) Kerr and (c) Faraday angles and (b) Kerr and (d) Faraday ellipticity as functions of photon energy $\omega$ for bilayer AFM insulator MnBi$_2$Te$_4$ with external magnetic moment $m_s$. The arrows labeled by $T_{i=1,2,3}$ correspond to the optical interband transitions as shown in Fig.~\ref{fig:layerFIM1}. The values of $\theta_K$ and $\theta_F$ in the low-energy limit are highlighted in (a) and (b). Parameters: $m=0.12$, $m_s=0.18$, $\epsilon_F=0$, $\epsilon_r=4$~\cite{tse2011,catarina2020}, $\mu_r=1$, $\hbar/\tau_s=0.02$.}
\label{fig:layerFIM2}
\end{figure}

The Kerr and Faraday angles and ellipticity are given by Fig.~\ref{fig:layerFIM2}. For low photon energy $\omega$, the magnitude of $\theta_{K/F}$ and $\gamma_{K/F}$ for AFMTI ($M_0=0.4$) is much larger than trivial insulator ($M_0=-0.2$). For $M_0=0.4$, the peaks (jumps) of $\theta_{K/F}$ ($\gamma_{K/F}$) arise from interband transitions $T_{i=1,2}$ at the $\Gamma$ point of Brillouin zone (see Fig.~\ref{fig:layerFIM1}). Note that there are some interband transitions missing between $T_1$ and $T_2$ since their interband velocity matrix elements vanish as a result of symmetries at the $\Gamma$ point. Therefore by proximity to magnetic substrates, we find another way to distinguish between AFMTI and trivial insulator.

\begin{figure}[htbp]
\centering \includegraphics[width=0.49\textwidth]{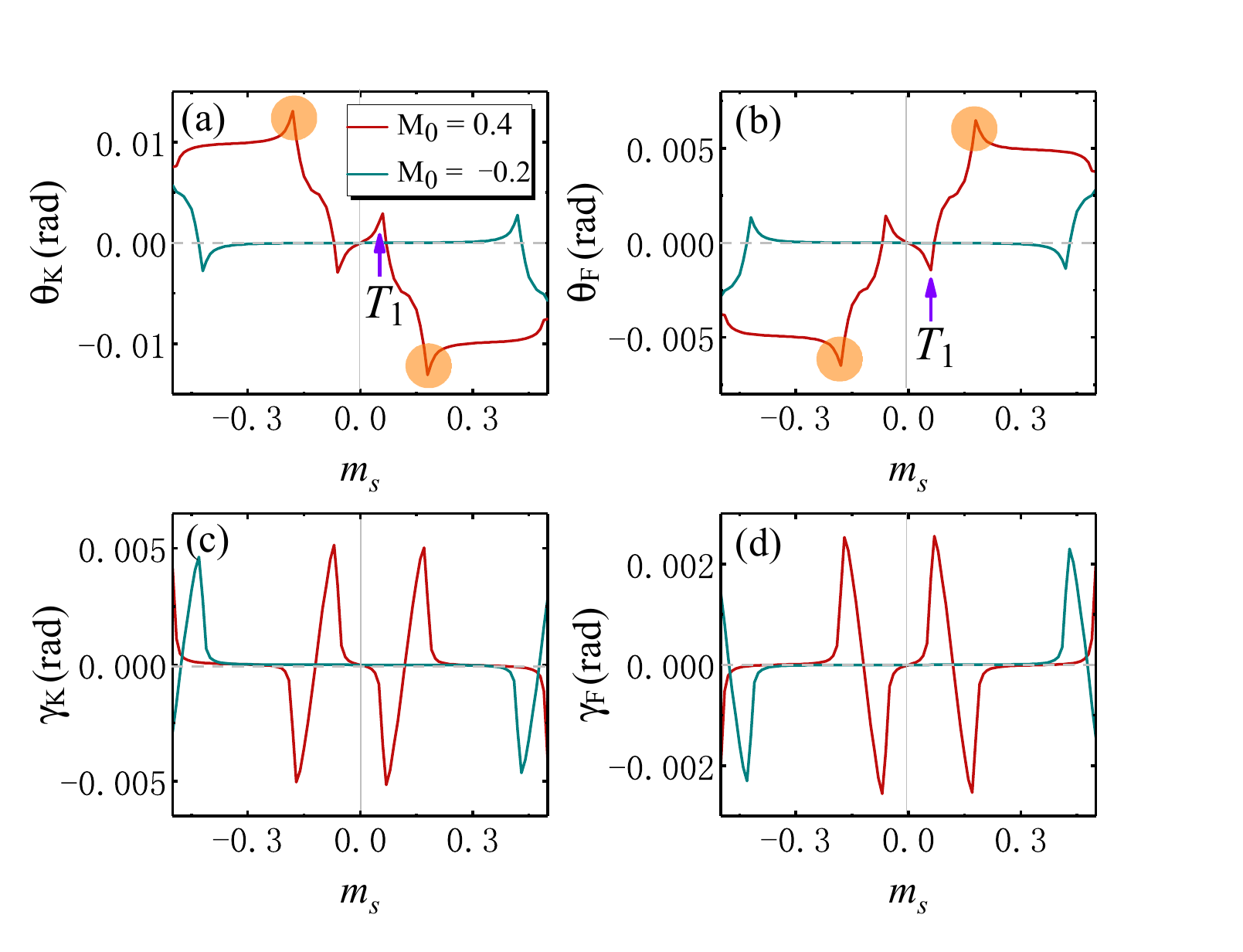}
\caption{ (a) Kerr and (c) Faraday angles and (b) Kerr and (d) Faraday ellipticity as functions of external magnetic moment $m_s$ for bilayer AFM insulator MnBi$_2$Te$_4$. Red and green lines correspond to AFMTI ($M_0=0.4$) and trivial insulator ($M_0=-0.2$), respectively. The peaks highlighted in orange region are the critical points satisfying $|m_s|=m$. Parameters: $m=0.12$, $\omega=0.12$, $\epsilon_F=0$, $\epsilon_r=4$~\cite{tse2011,catarina2020}, $\mu_r=1$, $\hbar/\tau_s=0.02$.}
\label{fig:layerFIM3}
\end{figure}

Magnetic moment $m_s$ dependence of Kerr and Faraday angles and ellipticity is plotted in Fig.~\ref{fig:layerFIM3}, where $\theta_{K/F}$ and $\gamma_{K/F}$ flip signs under the inversion of magnetic moment $m_s$. For AFMTI ($M_0=0.4$), $\theta_{K/F}$ reaches peaks at the $T_1$ transition, then it increases rapidly and almost saturates after some critical points. The critical points highlighted in the figure are given by the conditions $|m_s|=m$, which represents a change in the spin configuration. Physically, when $|m_s|<m$, the net magnetization at each layer is opposite, whereas when $|m_s|>m$, the net magnetization becomes the same. By contrast, such change of spin configurations does not lead to any qualitative differences in $\theta_{K/F}$ or $\gamma_{K/F}$ for trivial insulators. This provides another differences between AFMTI and trivial insulators.

\section{\label{sec:conclusion}Conclusions and discussions}

To summarize, we show that AFMTI and trivial insulators are distinguishable in magneto-optical Kerr and Faraday effects. To allow this, all mirror or combined mirror symmetries must be broken. We propose two feasible ways to break these symmetries, that is, by adding perpendicular electric field or external magnetic moment. In the former case, Kerr and Faraday angles and ellipticity show qualitatively different behaviors between AFMTI and trivial insulators, as a result of topological phase transitions by tuning parameters $M_0$. The significant differences of optical responses across topological phase transition have also been proposed in group-IV monochalcogenides SnSe~\cite{zhou2020}. The latter situation can be realized by proximity to magnetic substrate. In that case, Kerr and Faraday angles can provide the crucial information of spin configurations of the AFMTI phase. Our studies can be further generalized to multilayer AFM case, to investigate the role of layer numbers in magneto-optical responses.

Not only for bilayer AFM insulators MnBi$_2$Te$_4$, our theory can be extended to other $\mathcal{PT}$-symmetric AFM materials. In a recent paper, in-plane anomalous Hall effect has been studied in the general $\mathcal{PT}$-symmetric AFM materials~\cite{cao2023}. Given that Kerr and Faraday angles are closely related to the anomalous Hall conductivity in the low-frequency limit, anomalous longitudinal or transverse Kerr and Faraday effect, i.e., the magnetization is in-plane, may arise when symmetry conditions are satisfied. This leads to qualitatively different physics from the polar Kerr and Faraday effect studied in our work. Detailed analysis of this issue will be the object of future work.

\section*{ Acknowledgement}

This work is supported by the National Natural Science Foundation of China (NSFC, Grant No. 11904062), the Starting Research Fund from Guangzhou University (Grant No. RQ2020076) and Guangzhou Basic Research Program, jointed funded by Guangzhou University (Grant No. 202201020186).

\appendix

\section{\label{sec:sym}Symmetry analysis}

When $V=m_s=0$, the Hamiltonian $h(\bm k)=\bigoplus_{z=1}^{n_z}[h_a(\bm k,z)+h_b(\bm k,z)]$ preserves the symmetries $\mathcal{M}_{1/2}$, that is, $h(\bm k)\mathcal{M}_{1/2}=\mathcal{M}_{1/2}h(\bm k)$. $\epsilon_{\bm k\mu}$ and $|\bm k,\mu\rangle$ are eigenvalues and eigenstates of Hamiltonian $h(\bm k)$. Due to the presence of Kramer's degeneracy, we have $\epsilon_{\bm k\mu}=\epsilon_{\bm k\bar{\mu}}$, where $\mu(\bar{\mu})$ labels a pair of degenerate states. $\mathcal{M}_{1/2}$ symmetry relates the degenerate states between $\bm k$ and $\mathcal{M}_{1}\bm k=(-k_x,k_y)$ or $\mathcal{M}_{2}\bm k=(k_x,-k_y)$. For example, for $\mathcal{M}_1$, we have
\begin{equation}
\begin{split}
\left[\begin{array}{cc}
|\mathcal{M}_1\bm k,\mu\rangle & |\mathcal{M}_1\bm k,\bar{\mu}\rangle \\
\end{array}\right]=\mathcal{M}_1
\left[\begin{array}{cc}
|\bm k,\mu\rangle & |\bm k,\bar{\mu}\rangle \\
\end{array}\right]U,
\end{split}
\end{equation}
where $U$ is arbitrary $U(2)$ rotation. By grouping these states into Kramer's pairs, the conductivity tensor $\sigma_{\alpha\beta}$ from Eq. (\ref{conductivity}) can be rewritten as
\begin{equation}
\begin{split}
&\sigma_{xx}(\omega)=i\hbar\sum_{\epsilon_{\mu(\bar{\mu})}>0,\epsilon_{\mu'(\bar{\mu})}<0} \frac{\mathrm{Tr}[B_x(\bm k)B_x^+(\bm k)]}{\epsilon _{\bm k\mu'}-\epsilon_{\bm k \mu}}\\
&\times[\frac{1}{\omega+\epsilon_{\bm k\mu}-\epsilon _{\bm k\mu'}+i\hbar/2\tau_s}
+\frac{1}{\omega+\epsilon_{\bm k\mu'}-\epsilon _{\bm k\mu}+i\hbar/2\tau_s}],
\end{split}
\end{equation}

\begin{equation}
\begin{split}
&\sigma_{xy}(\omega)=i\hbar\sum_{\epsilon_{\mu(\bar{\mu})}>0,\epsilon_{\mu'(\bar{\mu})}<0} \frac{1}{\epsilon_{\bm k \mu'}-\epsilon _{\bm k\mu}}\\
&\times[\frac{\mathrm{Tr}[B_x(\bm k)B_y^+(\bm k)]}{\omega+\epsilon_{\bm k\mu}-\epsilon _{\bm k\mu'}+i\hbar/2\tau_s}
+\frac{\mathrm{Tr}[B_y(\bm k)B_x^+(\bm k)]}{\omega+\epsilon_{\bm k\mu'}-\epsilon _{\bm k\mu}+i\hbar/2\tau_s}],
\end{split}
\end{equation}
where the current matrix
\begin{equation}
\begin{split}
B_{x/y}(\bm k)=\left[\begin{array}{cc}
\langle\bm k,\mu| \\
\langle\bm k,\bar{\mu}| \\
\end{array}\right]j_{x/y}\left[\begin{array}{cc}
|\bm k,\mu'\rangle &  |\bm k,\bar{\mu}'\rangle  \\
\end{array}\right].
\end{split}
\end{equation}
For $\mathcal{M}_1$ symmetry,
\begin{equation}
\begin{split}
&B_x(\mathcal{M}_1\bm k)\\
&=\frac{e}{\hbar}\left[\begin{array}{cc}
\langle\mathcal{M}_1\bm k,\mu| \\
\langle\mathcal{M}_1\bm k,\bar{\mu}| \\
\end{array}\right]\frac{\partial h(\mathcal{M}_1\bm k)}{\partial(\mathcal{M}_1\bm k)_x}\left[\begin{array}{cc}
|\mathcal{M}_1\bm k,\mu'\rangle &  |\mathcal{M}_1\bm k,\bar{\mu}'\rangle  \\
\end{array}\right]\\
&=\frac{e}{\hbar}U^+\left[\begin{array}{cc}
\langle\bm k,\mu| \\
\langle\bm k,\bar{\mu}| \\
\end{array}\right]\mathcal{M}_1^+\frac{\partial h(\mathcal{M}_1\bm k)}{\partial(\mathcal{M}_1\bm k)_x}\mathcal{M}_1
\left[\begin{array}{cc}
|\bm k,\mu'\rangle & |\bm k,\bar{\mu}'\rangle \\
\end{array}\right]U\\
&=-U^+B_x(\bm k)U.
\end{split}
\end{equation}
Similarly, we have $B_y(\mathcal{M}_1\bm k)=U^+B_y(\bm k)U$. This means that the contributions to $\sigma_{xy}(\omega)$ from $\bm k$ and $\mathcal{M}_1\bm k$ always cancel with each other, resulting in $\sigma_{xy}(\omega)=0$. Similar discussion can be carried out for $\mathcal{M}_2$ symmetry, which also leads to zero Hall conductivity. When a perpendicular electric field or external magnetic moment is introduced, i.e., $V\neq0$ or $m_s\neq0$, $\mathcal{M}_{1/2}$ symmetries are broken, thus $\sigma_{xy}(\omega)$ can be nonzero.


\begin{thebibliography}{45}%
\makeatletter
\providecommand \@ifxundefined [1]{%
 \@ifx{#1\undefined}
}%
\providecommand \@ifnum [1]{%
 \ifnum #1\expandafter \@firstoftwo
 \else \expandafter \@secondoftwo
 \fi
}%
\providecommand \@ifx [1]{%
 \ifx #1\expandafter \@firstoftwo
 \else \expandafter \@secondoftwo
 \fi
}%
\providecommand \natexlab [1]{#1}%
\providecommand \enquote  [1]{``#1''}%
\providecommand \bibnamefont  [1]{#1}%
\providecommand \bibfnamefont [1]{#1}%
\providecommand \citenamefont [1]{#1}%
\providecommand \href@noop [0]{\@secondoftwo}%
\providecommand \href [0]{\begingroup \@sanitize@url \@href}%
\providecommand \@href[1]{\@@startlink{#1}\@@href}%
\providecommand \@@href[1]{\endgroup#1\@@endlink}%
\providecommand \@sanitize@url [0]{\catcode `\\12\catcode `\$12\catcode
  `\&12\catcode `\#12\catcode `\^12\catcode `\_12\catcode `\%12\relax}%
\providecommand \@@startlink[1]{}%
\providecommand \@@endlink[0]{}%
\providecommand \url  [0]{\begingroup\@sanitize@url \@url }%
\providecommand \@url [1]{\endgroup\@href {#1}{\urlprefix }}%
\providecommand \urlprefix  [0]{URL }%
\providecommand \Eprint [0]{\href }%
\providecommand \doibase [0]{http://dx.doi.org/}%
\providecommand \selectlanguage [0]{\@gobble}%
\providecommand \bibinfo  [0]{\@secondoftwo}%
\providecommand \bibfield  [0]{\@secondoftwo}%
\providecommand \translation [1]{[#1]}%
\providecommand \BibitemOpen [0]{}%
\providecommand \bibitemStop [0]{}%
\providecommand \bibitemNoStop [0]{.\EOS\space}%
\providecommand \EOS [0]{\spacefactor3000\relax}%
\providecommand \BibitemShut  [1]{\csname bibitem#1\endcsname}%
\let\auto@bib@innerbib\@empty
\bibitem [{\citenamefont {Jungwirth}\ \emph {et~al.}(2016)\citenamefont
  {Jungwirth}, \citenamefont {Marti}, \citenamefont {Wadley},\ and\
  \citenamefont {Wunderlich}}]{jungwirth2016}%
  \BibitemOpen
  \bibfield  {author} {\bibinfo {author} {\bibfnamefont {T.}~\bibnamefont
  {Jungwirth}}, \bibinfo {author} {\bibfnamefont {X.}~\bibnamefont {Marti}},
  \bibinfo {author} {\bibfnamefont {P.}~\bibnamefont {Wadley}}, \ and\ \bibinfo
  {author} {\bibfnamefont {J.}~\bibnamefont {Wunderlich}},\ }\href {\doibase
  10.1038/nnano.2016.18} {\bibfield  {journal} {\bibinfo  {journal} {Nat.
  Nanotech.}\ }\textbf {\bibinfo {volume} {11}},\ \bibinfo {pages} {231}
  (\bibinfo {year} {2016})}\BibitemShut {NoStop}%
\bibitem [{\citenamefont {Baltz}\ \emph {et~al.}(2018)\citenamefont {Baltz},
  \citenamefont {Manchon}, \citenamefont {Tsoi}, \citenamefont {Moriyama},
  \citenamefont {Ono},\ and\ \citenamefont {Tserkovnyak}}]{baltz2018}%
  \BibitemOpen
  \bibfield  {author} {\bibinfo {author} {\bibfnamefont {V.}~\bibnamefont
  {Baltz}}, \bibinfo {author} {\bibfnamefont {A.}~\bibnamefont {Manchon}},
  \bibinfo {author} {\bibfnamefont {M.}~\bibnamefont {Tsoi}}, \bibinfo {author}
  {\bibfnamefont {T.}~\bibnamefont {Moriyama}}, \bibinfo {author}
  {\bibfnamefont {T.}~\bibnamefont {Ono}}, \ and\ \bibinfo {author}
  {\bibfnamefont {Y.}~\bibnamefont {Tserkovnyak}},\ }\href {\doibase
  10.1103/RevModPhys.90.015005} {\bibfield  {journal} {\bibinfo  {journal}
  {Rev. Mod. Phys.}\ }\textbf {\bibinfo {volume} {90}},\ \bibinfo {pages}
  {015005} (\bibinfo {year} {2018})}\BibitemShut {NoStop}%
\bibitem [{\citenamefont {Li}\ \emph {et~al.}(2010)\citenamefont {Li},
  \citenamefont {Wang}, \citenamefont {Qi},\ and\ \citenamefont
  {Zhang}}]{li2010}%
  \BibitemOpen
  \bibfield  {author} {\bibinfo {author} {\bibfnamefont {R.}~\bibnamefont
  {Li}}, \bibinfo {author} {\bibfnamefont {J.}~\bibnamefont {Wang}}, \bibinfo
  {author} {\bibfnamefont {X.-L.}\ \bibnamefont {Qi}}, \ and\ \bibinfo {author}
  {\bibfnamefont {S.-C.}\ \bibnamefont {Zhang}},\ }\href {\doibase
  10.1038/nphys1534} {\bibfield  {journal} {\bibinfo  {journal} {Nat. Phys.}\
  }\textbf {\bibinfo {volume} {6}},\ \bibinfo {pages} {284} (\bibinfo {year}
  {2010})}\BibitemShut {NoStop}%
\bibitem [{\citenamefont {Nenno}\ \emph {et~al.}(2020)\citenamefont {Nenno},
  \citenamefont {Garcia}, \citenamefont {Gooth}, \citenamefont {Felser},\ and\
  \citenamefont {Narang}}]{nenno2020}%
  \BibitemOpen
  \bibfield  {author} {\bibinfo {author} {\bibfnamefont {D.~M.}\ \bibnamefont
  {Nenno}}, \bibinfo {author} {\bibfnamefont {C.~A.~C.}\ \bibnamefont
  {Garcia}}, \bibinfo {author} {\bibfnamefont {J.}~\bibnamefont {Gooth}},
  \bibinfo {author} {\bibfnamefont {C.}~\bibnamefont {Felser}}, \ and\ \bibinfo
  {author} {\bibfnamefont {P.}~\bibnamefont {Narang}},\ }\href {\doibase
  10.1038/s42254-020-0240-2} {\bibfield  {journal} {\bibinfo  {journal} {Nat.
  Rev. Phys.}\ }\textbf {\bibinfo {volume} {2}},\ \bibinfo {pages} {682}
  (\bibinfo {year} {2020})}\BibitemShut {NoStop}%
\bibitem [{\citenamefont {Liu}\ \emph {et~al.}(2020)\citenamefont {Liu},
  \citenamefont {Wang}, \citenamefont {Li}, \citenamefont {Wu}, \citenamefont
  {Li}, \citenamefont {Li}, \citenamefont {He}, \citenamefont {Xu},
  \citenamefont {Zhang},\ and\ \citenamefont {Wang}}]{liu2020}%
  \BibitemOpen
  \bibfield  {author} {\bibinfo {author} {\bibfnamefont {C.}~\bibnamefont
  {Liu}}, \bibinfo {author} {\bibfnamefont {Y.}~\bibnamefont {Wang}}, \bibinfo
  {author} {\bibfnamefont {H.}~\bibnamefont {Li}}, \bibinfo {author}
  {\bibfnamefont {Y.}~\bibnamefont {Wu}}, \bibinfo {author} {\bibfnamefont
  {Y.}~\bibnamefont {Li}}, \bibinfo {author} {\bibfnamefont {J.}~\bibnamefont
  {Li}}, \bibinfo {author} {\bibfnamefont {K.}~\bibnamefont {He}}, \bibinfo
  {author} {\bibfnamefont {Y.}~\bibnamefont {Xu}}, \bibinfo {author}
  {\bibfnamefont {J.}~\bibnamefont {Zhang}}, \ and\ \bibinfo {author}
  {\bibfnamefont {Y.}~\bibnamefont {Wang}},\ }\href {\doibase
  10.1038/s41563-019-0573-3} {\bibfield  {journal} {\bibinfo  {journal} {Nat.
  Mater.}\ }\textbf {\bibinfo {volume} {19}},\ \bibinfo {pages} {522} (\bibinfo
  {year} {2020})}\BibitemShut {NoStop}%
\bibitem [{\citenamefont {Essin}\ \emph {et~al.}(2009)\citenamefont {Essin},
  \citenamefont {Moore},\ and\ \citenamefont {Vanderbilt}}]{essin2009}%
  \BibitemOpen
  \bibfield  {author} {\bibinfo {author} {\bibfnamefont {A.~M.}\ \bibnamefont
  {Essin}}, \bibinfo {author} {\bibfnamefont {J.~E.}\ \bibnamefont {Moore}}, \
  and\ \bibinfo {author} {\bibfnamefont {D.}~\bibnamefont {Vanderbilt}},\
  }\href {\doibase 10.1103/PhysRevLett.102.146805} {\bibfield  {journal}
  {\bibinfo  {journal} {Phys. Rev. Lett.}\ }\textbf {\bibinfo {volume} {102}},\
  \bibinfo {pages} {146805} (\bibinfo {year} {2009})}\BibitemShut {NoStop}%
\bibitem [{\citenamefont {Zhang}\ \emph {et~al.}(2019)\citenamefont {Zhang},
  \citenamefont {Shi}, \citenamefont {Zhu}, \citenamefont {Xing}, \citenamefont
  {Zhang},\ and\ \citenamefont {Wang}}]{zhang2019}%
  \BibitemOpen
  \bibfield  {author} {\bibinfo {author} {\bibfnamefont {D.}~\bibnamefont
  {Zhang}}, \bibinfo {author} {\bibfnamefont {M.}~\bibnamefont {Shi}}, \bibinfo
  {author} {\bibfnamefont {T.}~\bibnamefont {Zhu}}, \bibinfo {author}
  {\bibfnamefont {D.}~\bibnamefont {Xing}}, \bibinfo {author} {\bibfnamefont
  {H.}~\bibnamefont {Zhang}}, \ and\ \bibinfo {author} {\bibfnamefont
  {J.}~\bibnamefont {Wang}},\ }\href {\doibase 10.1103/PhysRevLett.122.206401}
  {\bibfield  {journal} {\bibinfo  {journal} {Phys. Rev. Lett.}\ }\textbf
  {\bibinfo {volume} {122}},\ \bibinfo {pages} {206401} (\bibinfo {year}
  {2019})}\BibitemShut {NoStop}%
\bibitem [{\citenamefont {Armitage}\ and\ \citenamefont
  {Wu}(2019)}]{armitage2019}%
  \BibitemOpen
  \bibfield  {author} {\bibinfo {author} {\bibfnamefont {N.~P.}\ \bibnamefont
  {Armitage}}\ and\ \bibinfo {author} {\bibfnamefont {L.}~\bibnamefont {Wu}},\
  }\href {\doibase 10.21468/SciPostPhys.6.4.046} {\bibfield  {journal}
  {\bibinfo  {journal} {SciPost Phys.}\ }\textbf {\bibinfo {volume} {6}},\
  \bibinfo {pages} {046} (\bibinfo {year} {2019})}\BibitemShut {NoStop}%
\bibitem [{\citenamefont {Chen}\ \emph {et~al.}(2014)\citenamefont {Chen},
  \citenamefont {Niu},\ and\ \citenamefont {MacDonald}}]{chen2014}%
  \BibitemOpen
  \bibfield  {author} {\bibinfo {author} {\bibfnamefont {H.}~\bibnamefont
  {Chen}}, \bibinfo {author} {\bibfnamefont {Q.}~\bibnamefont {Niu}}, \ and\
  \bibinfo {author} {\bibfnamefont {A.~H.}\ \bibnamefont {MacDonald}},\ }\href
  {\doibase 10.1103/PhysRevLett.112.017205} {\bibfield  {journal} {\bibinfo
  {journal} {Phys. Rev. Lett.}\ }\textbf {\bibinfo {volume} {112}},\ \bibinfo
  {pages} {017205} (\bibinfo {year} {2014})}\BibitemShut {NoStop}%
\bibitem [{\citenamefont {Nakatsuji}\ \emph {et~al.}(2015)\citenamefont
  {Nakatsuji}, \citenamefont {Kiyohara},\ and\ \citenamefont
  {Higo}}]{nakatsuji2015}%
  \BibitemOpen
  \bibfield  {author} {\bibinfo {author} {\bibfnamefont {S.}~\bibnamefont
  {Nakatsuji}}, \bibinfo {author} {\bibfnamefont {N.}~\bibnamefont {Kiyohara}},
  \ and\ \bibinfo {author} {\bibfnamefont {T.}~\bibnamefont {Higo}},\ }\href
  {\doibase 10.1038/nature15723} {\bibfield  {journal} {\bibinfo  {journal}
  {Nature}\ }\textbf {\bibinfo {volume} {527}},\ \bibinfo {pages} {212}
  (\bibinfo {year} {2015})}\BibitemShut {NoStop}%
\bibitem [{\citenamefont {Nayak}\ \emph {et~al.}(2016)\citenamefont {Nayak},
  \citenamefont {Fischer}, \citenamefont {Sun}, \citenamefont {Yan},
  \citenamefont {Karel}, \citenamefont {Komarek}, \citenamefont {Shekhar},
  \citenamefont {Kumar}, \citenamefont {Schnelle}, \citenamefont {K\"{u}bler},
  \citenamefont {Felser},\ and\ \citenamefont {Parkin}}]{nayak2016}%
  \BibitemOpen
  \bibfield  {author} {\bibinfo {author} {\bibfnamefont {A.~A.}\ \bibnamefont
  {Nayak}}, \bibinfo {author} {\bibfnamefont {J.~E.}\ \bibnamefont {Fischer}},
  \bibinfo {author} {\bibfnamefont {Y.}~\bibnamefont {Sun}}, \bibinfo {author}
  {\bibfnamefont {B.}~\bibnamefont {Yan}}, \bibinfo {author} {\bibfnamefont
  {J.}~\bibnamefont {Karel}}, \bibinfo {author} {\bibfnamefont {A.~C.}\
  \bibnamefont {Komarek}}, \bibinfo {author} {\bibfnamefont {C.}~\bibnamefont
  {Shekhar}}, \bibinfo {author} {\bibfnamefont {N.}~\bibnamefont {Kumar}},
  \bibinfo {author} {\bibfnamefont {W.}~\bibnamefont {Schnelle}}, \bibinfo
  {author} {\bibfnamefont {J.}~\bibnamefont {K\"{u}bler}}, \bibinfo {author}
  {\bibfnamefont {C.}~\bibnamefont {Felser}}, \ and\ \bibinfo {author}
  {\bibfnamefont {S.~S.~P.}\ \bibnamefont {Parkin}},\ }\href {\doibase
  10.1126/sciadv.1501870} {\bibfield  {journal} {\bibinfo  {journal} {Science
  Advances}\ }\textbf {\bibinfo {volume} {2}},\ \bibinfo {pages} {e1501870}
  (\bibinfo {year} {2016})}\BibitemShut {NoStop}%
\bibitem [{\citenamefont {Feng}\ \emph {et~al.}(2020)\citenamefont {Feng},
  \citenamefont {Hanke}, \citenamefont {Zhou}, \citenamefont {Guo},
  \citenamefont {Blugel}, \citenamefont {Mokrousov},\ and\ \citenamefont
  {Yao}}]{feng2020}%
  \BibitemOpen
  \bibfield  {author} {\bibinfo {author} {\bibfnamefont {W.}~\bibnamefont
  {Feng}}, \bibinfo {author} {\bibfnamefont {J.-P.}\ \bibnamefont {Hanke}},
  \bibinfo {author} {\bibfnamefont {X.}~\bibnamefont {Zhou}}, \bibinfo {author}
  {\bibfnamefont {G.-Y.}\ \bibnamefont {Guo}}, \bibinfo {author} {\bibfnamefont
  {S.}~\bibnamefont {Blugel}}, \bibinfo {author} {\bibfnamefont
  {Y.}~\bibnamefont {Mokrousov}}, \ and\ \bibinfo {author} {\bibfnamefont
  {Y.}~\bibnamefont {Yao}},\ }\href {\doibase 10.1038/s41467-019-13968-8}
  {\bibfield  {journal} {\bibinfo  {journal} {Nat. Commun.}\ }\textbf {\bibinfo
  {volume} {11}},\ \bibinfo {pages} {118} (\bibinfo {year} {2020})}\BibitemShut
  {NoStop}%
\bibitem [{\citenamefont {Otrokov}\ \emph
  {et~al.}(2019{\natexlab{a}})\citenamefont {Otrokov}, \citenamefont
  {Klimovskikh}, \citenamefont {Bentmann}, \citenamefont {Estyunin},
  \citenamefont {Zeugner}, \citenamefont {Aliev}, \citenamefont {Gaß},
  \citenamefont {Wolter}, \citenamefont {Koroleva}, \citenamefont {Shikin},
  \citenamefont {Blanco-Rey}, \citenamefont {Hoffmann}, \citenamefont
  {Rusinov}, \citenamefont {Vyazovskaya}, \citenamefont {Eremeev},
  \citenamefont {Koroteev}, \citenamefont {Kuznetsov}, \citenamefont {Freyse},
  \citenamefont {S\'{a}nchez-Barriga}, \citenamefont {Amiraslanov},
  \citenamefont {Babanly}, \citenamefont {Mamedov}, \citenamefont {Abdullayev},
  \citenamefont {Zverev}, \citenamefont {Alfonsov}, \citenamefont {Kataev},
  \citenamefont {B\"{u}chner}, \citenamefont {Schwier}, \citenamefont {Kumar},
  \citenamefont {Kimura}, \citenamefont {Petaccia}, \citenamefont {Di~Santo},
  \citenamefont {Vidal}, \citenamefont {Schatz}, \citenamefont {Kißner},
  \citenamefont {\"{U}nzelmann}, \citenamefont {Min}, \citenamefont {Moser},
  \citenamefont {Peixoto}, \citenamefont {Reinert}, \citenamefont {Ernst},
  \citenamefont {Echenique}, \citenamefont {Isaeva},\ and\ \citenamefont
  {Chulkov}}]{otrokov2019}%
  \BibitemOpen
  \bibfield  {author} {\bibinfo {author} {\bibfnamefont {M.~M.}\ \bibnamefont
  {Otrokov}}, \bibinfo {author} {\bibfnamefont {I.~I.}\ \bibnamefont
  {Klimovskikh}}, \bibinfo {author} {\bibfnamefont {H.}~\bibnamefont
  {Bentmann}}, \bibinfo {author} {\bibfnamefont {D.}~\bibnamefont {Estyunin}},
  \bibinfo {author} {\bibfnamefont {A.}~\bibnamefont {Zeugner}}, \bibinfo
  {author} {\bibfnamefont {Z.~S.}\ \bibnamefont {Aliev}}, \bibinfo {author}
  {\bibfnamefont {S.}~\bibnamefont {Gaß}}, \bibinfo {author} {\bibfnamefont
  {A.~U.~B.}\ \bibnamefont {Wolter}}, \bibinfo {author} {\bibfnamefont {A.~V.}\
  \bibnamefont {Koroleva}}, \bibinfo {author} {\bibfnamefont {A.~M.}\
  \bibnamefont {Shikin}}, \bibinfo {author} {\bibfnamefont {M.}~\bibnamefont
  {Blanco-Rey}}, \bibinfo {author} {\bibfnamefont {M.}~\bibnamefont
  {Hoffmann}}, \bibinfo {author} {\bibfnamefont {I.~P.}\ \bibnamefont
  {Rusinov}}, \bibinfo {author} {\bibfnamefont {A.~Y.}\ \bibnamefont
  {Vyazovskaya}}, \bibinfo {author} {\bibfnamefont {S.~V.}\ \bibnamefont
  {Eremeev}}, \bibinfo {author} {\bibfnamefont {Y.~M.}\ \bibnamefont
  {Koroteev}}, \bibinfo {author} {\bibfnamefont {V.~M.}\ \bibnamefont
  {Kuznetsov}}, \bibinfo {author} {\bibfnamefont {F.}~\bibnamefont {Freyse}},
  \bibinfo {author} {\bibfnamefont {J.}~\bibnamefont {S\'{a}nchez-Barriga}},
  \bibinfo {author} {\bibfnamefont {I.~R.}\ \bibnamefont {Amiraslanov}},
  \bibinfo {author} {\bibfnamefont {M.~B.}\ \bibnamefont {Babanly}}, \bibinfo
  {author} {\bibfnamefont {N.~T.}\ \bibnamefont {Mamedov}}, \bibinfo {author}
  {\bibfnamefont {N.~A.}\ \bibnamefont {Abdullayev}}, \bibinfo {author}
  {\bibfnamefont {V.~N.}\ \bibnamefont {Zverev}}, \bibinfo {author}
  {\bibfnamefont {A.}~\bibnamefont {Alfonsov}}, \bibinfo {author}
  {\bibfnamefont {V.}~\bibnamefont {Kataev}}, \bibinfo {author} {\bibfnamefont
  {B.}~\bibnamefont {B\"{u}chner}}, \bibinfo {author} {\bibfnamefont {E.~F.}\
  \bibnamefont {Schwier}}, \bibinfo {author} {\bibfnamefont {S.}~\bibnamefont
  {Kumar}}, \bibinfo {author} {\bibfnamefont {A.}~\bibnamefont {Kimura}},
  \bibinfo {author} {\bibfnamefont {L.}~\bibnamefont {Petaccia}}, \bibinfo
  {author} {\bibfnamefont {G.}~\bibnamefont {Di~Santo}}, \bibinfo {author}
  {\bibfnamefont {R.~C.}\ \bibnamefont {Vidal}}, \bibinfo {author}
  {\bibfnamefont {S.}~\bibnamefont {Schatz}}, \bibinfo {author} {\bibfnamefont
  {K.}~\bibnamefont {Kißner}}, \bibinfo {author} {\bibfnamefont
  {M.}~\bibnamefont {\"{U}nzelmann}}, \bibinfo {author} {\bibfnamefont {C.~H.}\
  \bibnamefont {Min}}, \bibinfo {author} {\bibfnamefont {S.}~\bibnamefont
  {Moser}}, \bibinfo {author} {\bibfnamefont {T.~R.~F.}\ \bibnamefont
  {Peixoto}}, \bibinfo {author} {\bibfnamefont {F.}~\bibnamefont {Reinert}},
  \bibinfo {author} {\bibfnamefont {A.}~\bibnamefont {Ernst}}, \bibinfo
  {author} {\bibfnamefont {P.~M.}\ \bibnamefont {Echenique}}, \bibinfo {author}
  {\bibfnamefont {A.}~\bibnamefont {Isaeva}}, \ and\ \bibinfo {author}
  {\bibfnamefont {E.~V.}\ \bibnamefont {Chulkov}},\ }\href {\doibase
  10.1038/s41586-019-1840-9} {\bibfield  {journal} {\bibinfo  {journal}
  {Nature}\ }\textbf {\bibinfo {volume} {576}},\ \bibinfo {pages} {416}
  (\bibinfo {year} {2019}{\natexlab{a}})}\BibitemShut {NoStop}%
\bibitem [{\citenamefont {Otrokov}\ \emph
  {et~al.}(2019{\natexlab{b}})\citenamefont {Otrokov}, \citenamefont {Rusinov},
  \citenamefont {Blanco-Rey}, \citenamefont {Hoffmann}, \citenamefont
  {Vyazovskaya}, \citenamefont {Eremeev}, \citenamefont {Ernst}, \citenamefont
  {Echenique}, \citenamefont {Arnau},\ and\ \citenamefont
  {Chulkov}}]{otrokovm2019}%
  \BibitemOpen
  \bibfield  {author} {\bibinfo {author} {\bibfnamefont {M.~M.}\ \bibnamefont
  {Otrokov}}, \bibinfo {author} {\bibfnamefont {I.~P.}\ \bibnamefont
  {Rusinov}}, \bibinfo {author} {\bibfnamefont {M.}~\bibnamefont {Blanco-Rey}},
  \bibinfo {author} {\bibfnamefont {M.}~\bibnamefont {Hoffmann}}, \bibinfo
  {author} {\bibfnamefont {A.~Y.}\ \bibnamefont {Vyazovskaya}}, \bibinfo
  {author} {\bibfnamefont {S.~V.}\ \bibnamefont {Eremeev}}, \bibinfo {author}
  {\bibfnamefont {A.}~\bibnamefont {Ernst}}, \bibinfo {author} {\bibfnamefont
  {P.~M.}\ \bibnamefont {Echenique}}, \bibinfo {author} {\bibfnamefont
  {A.}~\bibnamefont {Arnau}}, \ and\ \bibinfo {author} {\bibfnamefont {E.~V.}\
  \bibnamefont {Chulkov}},\ }\href {\doibase 10.1103/PhysRevLett.122.107202}
  {\bibfield  {journal} {\bibinfo  {journal} {Phys. Rev. Lett.}\ }\textbf
  {\bibinfo {volume} {122}},\ \bibinfo {pages} {107202} (\bibinfo {year}
  {2019}{\natexlab{b}})}\BibitemShut {NoStop}%
\bibitem [{\citenamefont {Chen}\ \emph {et~al.}(2019)\citenamefont {Chen},
  \citenamefont {Xu}, \citenamefont {Li}, \citenamefont {Li}, \citenamefont
  {Wang}, \citenamefont {Zhang}, \citenamefont {Li}, \citenamefont {Wu},
  \citenamefont {Liang}, \citenamefont {Chen}, \citenamefont {Jung},
  \citenamefont {Cacho}, \citenamefont {Mao}, \citenamefont {Liu},
  \citenamefont {Wang}, \citenamefont {Guo}, \citenamefont {Xu}, \citenamefont
  {Liu}, \citenamefont {Yang},\ and\ \citenamefont {Chen}}]{chen2019}%
  \BibitemOpen
  \bibfield  {author} {\bibinfo {author} {\bibfnamefont {Y.~J.}\ \bibnamefont
  {Chen}}, \bibinfo {author} {\bibfnamefont {L.~X.}\ \bibnamefont {Xu}},
  \bibinfo {author} {\bibfnamefont {J.~H.}\ \bibnamefont {Li}}, \bibinfo
  {author} {\bibfnamefont {Y.~W.}\ \bibnamefont {Li}}, \bibinfo {author}
  {\bibfnamefont {H.~Y.}\ \bibnamefont {Wang}}, \bibinfo {author}
  {\bibfnamefont {C.~F.}\ \bibnamefont {Zhang}}, \bibinfo {author}
  {\bibfnamefont {H.}~\bibnamefont {Li}}, \bibinfo {author} {\bibfnamefont
  {Y.}~\bibnamefont {Wu}}, \bibinfo {author} {\bibfnamefont {A.~J.}\
  \bibnamefont {Liang}}, \bibinfo {author} {\bibfnamefont {C.}~\bibnamefont
  {Chen}}, \bibinfo {author} {\bibfnamefont {S.~W.}\ \bibnamefont {Jung}},
  \bibinfo {author} {\bibfnamefont {C.}~\bibnamefont {Cacho}}, \bibinfo
  {author} {\bibfnamefont {Y.~H.}\ \bibnamefont {Mao}}, \bibinfo {author}
  {\bibfnamefont {S.}~\bibnamefont {Liu}}, \bibinfo {author} {\bibfnamefont
  {M.~X.}\ \bibnamefont {Wang}}, \bibinfo {author} {\bibfnamefont {Y.~F.}\
  \bibnamefont {Guo}}, \bibinfo {author} {\bibfnamefont {Y.}~\bibnamefont
  {Xu}}, \bibinfo {author} {\bibfnamefont {Z.~K.}\ \bibnamefont {Liu}},
  \bibinfo {author} {\bibfnamefont {L.~X.}\ \bibnamefont {Yang}}, \ and\
  \bibinfo {author} {\bibfnamefont {Y.~L.}\ \bibnamefont {Chen}},\ }\href
  {\doibase 10.1103/PhysRevX.9.041040} {\bibfield  {journal} {\bibinfo
  {journal} {Phys. Rev. X}\ }\textbf {\bibinfo {volume} {9}},\ \bibinfo {pages}
  {041040} (\bibinfo {year} {2019})}\BibitemShut {NoStop}%
\bibitem [{\citenamefont {Hao}\ \emph {et~al.}(2019)\citenamefont {Hao},
  \citenamefont {Liu}, \citenamefont {Feng}, \citenamefont {Ma}, \citenamefont
  {Schwier}, \citenamefont {Arita}, \citenamefont {Kumar}, \citenamefont {Hu},
  \citenamefont {Lu}, \citenamefont {Zeng}, \citenamefont {Wang}, \citenamefont
  {Hao}, \citenamefont {Sun}, \citenamefont {Zhang}, \citenamefont {Mei},
  \citenamefont {Ni}, \citenamefont {Wu}, \citenamefont {Shimada},
  \citenamefont {Chen}, \citenamefont {Liu},\ and\ \citenamefont
  {Liu}}]{hao2019}%
  \BibitemOpen
  \bibfield  {author} {\bibinfo {author} {\bibfnamefont {Y.-J.}\ \bibnamefont
  {Hao}}, \bibinfo {author} {\bibfnamefont {P.}~\bibnamefont {Liu}}, \bibinfo
  {author} {\bibfnamefont {Y.}~\bibnamefont {Feng}}, \bibinfo {author}
  {\bibfnamefont {X.-M.}\ \bibnamefont {Ma}}, \bibinfo {author} {\bibfnamefont
  {E.~F.}\ \bibnamefont {Schwier}}, \bibinfo {author} {\bibfnamefont
  {M.}~\bibnamefont {Arita}}, \bibinfo {author} {\bibfnamefont
  {S.}~\bibnamefont {Kumar}}, \bibinfo {author} {\bibfnamefont
  {C.}~\bibnamefont {Hu}}, \bibinfo {author} {\bibfnamefont {R.}~\bibnamefont
  {Lu}}, \bibinfo {author} {\bibfnamefont {M.}~\bibnamefont {Zeng}}, \bibinfo
  {author} {\bibfnamefont {Y.}~\bibnamefont {Wang}}, \bibinfo {author}
  {\bibfnamefont {Z.}~\bibnamefont {Hao}}, \bibinfo {author} {\bibfnamefont
  {H.-Y.}\ \bibnamefont {Sun}}, \bibinfo {author} {\bibfnamefont
  {K.}~\bibnamefont {Zhang}}, \bibinfo {author} {\bibfnamefont
  {J.}~\bibnamefont {Mei}}, \bibinfo {author} {\bibfnamefont {N.}~\bibnamefont
  {Ni}}, \bibinfo {author} {\bibfnamefont {L.}~\bibnamefont {Wu}}, \bibinfo
  {author} {\bibfnamefont {K.}~\bibnamefont {Shimada}}, \bibinfo {author}
  {\bibfnamefont {C.}~\bibnamefont {Chen}}, \bibinfo {author} {\bibfnamefont
  {Q.}~\bibnamefont {Liu}}, \ and\ \bibinfo {author} {\bibfnamefont
  {C.}~\bibnamefont {Liu}},\ }\href {\doibase 10.1103/PhysRevX.9.041038}
  {\bibfield  {journal} {\bibinfo  {journal} {Phys. Rev. X}\ }\textbf {\bibinfo
  {volume} {9}},\ \bibinfo {pages} {041038} (\bibinfo {year}
  {2019})}\BibitemShut {NoStop}%
\bibitem [{\citenamefont {Swatek}\ \emph {et~al.}(2020)\citenamefont {Swatek},
  \citenamefont {Wu}, \citenamefont {Wang}, \citenamefont {Lee}, \citenamefont
  {Schrunk}, \citenamefont {Yan},\ and\ \citenamefont {Kaminski}}]{swatek2020}%
  \BibitemOpen
  \bibfield  {author} {\bibinfo {author} {\bibfnamefont {P.}~\bibnamefont
  {Swatek}}, \bibinfo {author} {\bibfnamefont {Y.}~\bibnamefont {Wu}}, \bibinfo
  {author} {\bibfnamefont {L.-L.}\ \bibnamefont {Wang}}, \bibinfo {author}
  {\bibfnamefont {K.}~\bibnamefont {Lee}}, \bibinfo {author} {\bibfnamefont
  {B.}~\bibnamefont {Schrunk}}, \bibinfo {author} {\bibfnamefont
  {J.}~\bibnamefont {Yan}}, \ and\ \bibinfo {author} {\bibfnamefont
  {A.}~\bibnamefont {Kaminski}},\ }\href {\doibase 10.1103/PhysRevB.101.161109}
  {\bibfield  {journal} {\bibinfo  {journal} {Phys. Rev. B}\ }\textbf {\bibinfo
  {volume} {101}},\ \bibinfo {pages} {161109} (\bibinfo {year}
  {2020})}\BibitemShut {NoStop}%
\bibitem [{\citenamefont {Zhang}\ \emph
  {et~al.}(2020{\natexlab{a}})\citenamefont {Zhang}, \citenamefont {Wang},
  \citenamefont {Shi}, \citenamefont {Zhu}, \citenamefont {Zhang},\ and\
  \citenamefont {Wang}}]{zhang2020}%
  \BibitemOpen
  \bibfield  {author} {\bibinfo {author} {\bibfnamefont {J.}~\bibnamefont
  {Zhang}}, \bibinfo {author} {\bibfnamefont {D.}~\bibnamefont {Wang}},
  \bibinfo {author} {\bibfnamefont {M.}~\bibnamefont {Shi}}, \bibinfo {author}
  {\bibfnamefont {T.}~\bibnamefont {Zhu}}, \bibinfo {author} {\bibfnamefont
  {H.}~\bibnamefont {Zhang}}, \ and\ \bibinfo {author} {\bibfnamefont
  {J.}~\bibnamefont {Wang}},\ }\href {\doibase 10.1088/0256-307X/37/7/077304}
  {\bibfield  {journal} {\bibinfo  {journal} {Chin. Phys. Lett.}\ }\textbf
  {\bibinfo {volume} {37}},\ \bibinfo {pages} {077304} (\bibinfo {year}
  {2020}{\natexlab{a}})}\BibitemShut {NoStop}%
\bibitem [{\citenamefont {Gao}\ \emph {et~al.}(2021)\citenamefont {Gao},
  \citenamefont {Liu}, \citenamefont {Hu}, \citenamefont {Qiu}, \citenamefont
  {Tzschaschel}, \citenamefont {Ghosh}, \citenamefont {Ho}, \citenamefont
  {B$\acute{e}$rub$\acute{e}$}, \citenamefont {Chen}, \citenamefont {Sun},
  \citenamefont {Zhang}, \citenamefont {Zhang}, \citenamefont {Wang},
  \citenamefont {Wang}, \citenamefont {Huang}, \citenamefont {Felser},
  \citenamefont {Agarwal}, \citenamefont {Ding}, \citenamefont {Tien},
  \citenamefont {Akey}, \citenamefont {Gardener}, \citenamefont {Singh},
  \citenamefont {Watanabe}, \citenamefont {Taniguchi}, \citenamefont {Burch},
  \citenamefont {Bell}, \citenamefont {Zhou}, \citenamefont {Gao},
  \citenamefont {Lu}, \citenamefont {Bansil}, \citenamefont {Lin},
  \citenamefont {Chang}, \citenamefont {Fu}, \citenamefont {Ma}, \citenamefont
  {Ni},\ and\ \citenamefont {Xu}}]{gao2021}%
  \BibitemOpen
  \bibfield  {author} {\bibinfo {author} {\bibfnamefont {A.}~\bibnamefont
  {Gao}}, \bibinfo {author} {\bibfnamefont {Y.-F.}\ \bibnamefont {Liu}},
  \bibinfo {author} {\bibfnamefont {C.}~\bibnamefont {Hu}}, \bibinfo {author}
  {\bibfnamefont {J.-X.}\ \bibnamefont {Qiu}}, \bibinfo {author} {\bibfnamefont
  {C.}~\bibnamefont {Tzschaschel}}, \bibinfo {author} {\bibfnamefont
  {B.}~\bibnamefont {Ghosh}}, \bibinfo {author} {\bibfnamefont {S.-C.}\
  \bibnamefont {Ho}}, \bibinfo {author} {\bibfnamefont {D.}~\bibnamefont
  {B$\acute{e}$rub$\acute{e}$}}, \bibinfo {author} {\bibfnamefont
  {R.}~\bibnamefont {Chen}}, \bibinfo {author} {\bibfnamefont {H.}~\bibnamefont
  {Sun}}, \bibinfo {author} {\bibfnamefont {Z.}~\bibnamefont {Zhang}}, \bibinfo
  {author} {\bibfnamefont {X.-Y.}\ \bibnamefont {Zhang}}, \bibinfo {author}
  {\bibfnamefont {Y.-X.}\ \bibnamefont {Wang}}, \bibinfo {author}
  {\bibfnamefont {N.}~\bibnamefont {Wang}}, \bibinfo {author} {\bibfnamefont
  {Z.}~\bibnamefont {Huang}}, \bibinfo {author} {\bibfnamefont
  {C.}~\bibnamefont {Felser}}, \bibinfo {author} {\bibfnamefont
  {A.}~\bibnamefont {Agarwal}}, \bibinfo {author} {\bibfnamefont
  {T.}~\bibnamefont {Ding}}, \bibinfo {author} {\bibfnamefont {H.-J.}\
  \bibnamefont {Tien}}, \bibinfo {author} {\bibfnamefont {A.}~\bibnamefont
  {Akey}}, \bibinfo {author} {\bibfnamefont {J.}~\bibnamefont {Gardener}},
  \bibinfo {author} {\bibfnamefont {B.}~\bibnamefont {Singh}}, \bibinfo
  {author} {\bibfnamefont {K.}~\bibnamefont {Watanabe}}, \bibinfo {author}
  {\bibfnamefont {T.}~\bibnamefont {Taniguchi}}, \bibinfo {author}
  {\bibfnamefont {K.~S.}\ \bibnamefont {Burch}}, \bibinfo {author}
  {\bibfnamefont {D.~C.}\ \bibnamefont {Bell}}, \bibinfo {author}
  {\bibfnamefont {B.~B.}\ \bibnamefont {Zhou}}, \bibinfo {author}
  {\bibfnamefont {W.}~\bibnamefont {Gao}}, \bibinfo {author} {\bibfnamefont
  {H.-Z.}\ \bibnamefont {Lu}}, \bibinfo {author} {\bibfnamefont
  {A.}~\bibnamefont {Bansil}}, \bibinfo {author} {\bibfnamefont
  {H.}~\bibnamefont {Lin}}, \bibinfo {author} {\bibfnamefont {T.-R.}\
  \bibnamefont {Chang}}, \bibinfo {author} {\bibfnamefont {L.}~\bibnamefont
  {Fu}}, \bibinfo {author} {\bibfnamefont {Q.}~\bibnamefont {Ma}}, \bibinfo
  {author} {\bibfnamefont {N.}~\bibnamefont {Ni}}, \ and\ \bibinfo {author}
  {\bibfnamefont {S.-Y.}\ \bibnamefont {Xu}},\ }\href {\doibase
  10.1038/s41586-021-03679-w} {\bibfield  {journal} {\bibinfo  {journal}
  {Nature}\ }\textbf {\bibinfo {volume} {595}},\ \bibinfo {pages} {521}
  (\bibinfo {year} {2021})}\BibitemShut {NoStop}%
\bibitem [{\citenamefont {Chen}\ \emph {et~al.}(2022)\citenamefont {Chen},
  \citenamefont {Sun}, \citenamefont {Gu}, \citenamefont {Hua}, \citenamefont
  {Liu}, \citenamefont {Lu},\ and\ \citenamefont {Xie}}]{chen2022}%
  \BibitemOpen
  \bibfield  {author} {\bibinfo {author} {\bibfnamefont {R.}~\bibnamefont
  {Chen}}, \bibinfo {author} {\bibfnamefont {H.-P.}\ \bibnamefont {Sun}},
  \bibinfo {author} {\bibfnamefont {M.}~\bibnamefont {Gu}}, \bibinfo {author}
  {\bibfnamefont {C.-B.}\ \bibnamefont {Hua}}, \bibinfo {author} {\bibfnamefont
  {Q.}~\bibnamefont {Liu}}, \bibinfo {author} {\bibfnamefont {H.-Z.}\
  \bibnamefont {Lu}}, \ and\ \bibinfo {author} {\bibfnamefont {X.~C.}\
  \bibnamefont {Xie}},\ }\href {\doibase 10.1093/nsr/nwac140} {\bibfield
  {journal} {\bibinfo  {journal} {National Science Review}\ ,\ \bibinfo {pages}
  {nwac140}} (\bibinfo {year} {2022})}\BibitemShut {NoStop}%
\bibitem [{\citenamefont {Chen}\ \emph {et~al.}(2021)\citenamefont {Chen},
  \citenamefont {Li}, \citenamefont {Sun}, \citenamefont {Liu}, \citenamefont
  {Zhao}, \citenamefont {Lu},\ and\ \citenamefont {Xie}}]{chen2021}%
  \BibitemOpen
  \bibfield  {author} {\bibinfo {author} {\bibfnamefont {R.}~\bibnamefont
  {Chen}}, \bibinfo {author} {\bibfnamefont {S.}~\bibnamefont {Li}}, \bibinfo
  {author} {\bibfnamefont {H.-P.}\ \bibnamefont {Sun}}, \bibinfo {author}
  {\bibfnamefont {Q.}~\bibnamefont {Liu}}, \bibinfo {author} {\bibfnamefont
  {Y.}~\bibnamefont {Zhao}}, \bibinfo {author} {\bibfnamefont {H.-Z.}\
  \bibnamefont {Lu}}, \ and\ \bibinfo {author} {\bibfnamefont {X.~C.}\
  \bibnamefont {Xie}},\ }\href {\doibase 10.1103/PhysRevB.103.L241409}
  {\bibfield  {journal} {\bibinfo  {journal} {Phys. Rev. B}\ }\textbf {\bibinfo
  {volume} {103}},\ \bibinfo {pages} {L241409} (\bibinfo {year}
  {2021})}\BibitemShut {NoStop}%
\bibitem [{\citenamefont {Ding}\ \emph {et~al.}(2020)\citenamefont {Ding},
  \citenamefont {Xu}, \citenamefont {Chen},\ and\ \citenamefont
  {Xie}}]{ding2020}%
  \BibitemOpen
  \bibfield  {author} {\bibinfo {author} {\bibfnamefont {Y.-R.}\ \bibnamefont
  {Ding}}, \bibinfo {author} {\bibfnamefont {D.-H.}\ \bibnamefont {Xu}},
  \bibinfo {author} {\bibfnamefont {C.-Z.}\ \bibnamefont {Chen}}, \ and\
  \bibinfo {author} {\bibfnamefont {X.~C.}\ \bibnamefont {Xie}},\ }\href
  {\doibase 10.1103/PhysRevB.101.041404} {\bibfield  {journal} {\bibinfo
  {journal} {Phys. Rev. B}\ }\textbf {\bibinfo {volume} {101}},\ \bibinfo
  {pages} {041404} (\bibinfo {year} {2020})}\BibitemShut {NoStop}%
\bibitem [{\citenamefont {Zhang}\ \emph
  {et~al.}(2020{\natexlab{b}})\citenamefont {Zhang}, \citenamefont {Wu},\ and\
  \citenamefont {Das~Sarma}}]{zhangr2020}%
  \BibitemOpen
  \bibfield  {author} {\bibinfo {author} {\bibfnamefont {R.-X.}\ \bibnamefont
  {Zhang}}, \bibinfo {author} {\bibfnamefont {F.}~\bibnamefont {Wu}}, \ and\
  \bibinfo {author} {\bibfnamefont {S.}~\bibnamefont {Das~Sarma}},\ }\href
  {\doibase 10.1103/PhysRevLett.124.136407} {\bibfield  {journal} {\bibinfo
  {journal} {Phys. Rev. Lett.}\ }\textbf {\bibinfo {volume} {124}},\ \bibinfo
  {pages} {136407} (\bibinfo {year} {2020}{\natexlab{b}})}\BibitemShut
  {NoStop}%
\bibitem [{\citenamefont {Li}\ \emph {et~al.}(2021)\citenamefont {Li},
  \citenamefont {Jiang}, \citenamefont {Chen},\ and\ \citenamefont
  {Xie}}]{li2021}%
  \BibitemOpen
  \bibfield  {author} {\bibinfo {author} {\bibfnamefont {H.}~\bibnamefont
  {Li}}, \bibinfo {author} {\bibfnamefont {H.}~\bibnamefont {Jiang}}, \bibinfo
  {author} {\bibfnamefont {C.-Z.}\ \bibnamefont {Chen}}, \ and\ \bibinfo
  {author} {\bibfnamefont {X.~C.}\ \bibnamefont {Xie}},\ }\href {\doibase
  10.1103/PhysRevLett.126.156601} {\bibfield  {journal} {\bibinfo  {journal}
  {Phys. Rev. Lett.}\ }\textbf {\bibinfo {volume} {126}},\ \bibinfo {pages}
  {156601} (\bibinfo {year} {2021})}\BibitemShut {NoStop}%
\bibitem [{\citenamefont {Gu}\ \emph {et~al.}(2021)\citenamefont {Gu},
  \citenamefont {Li}, \citenamefont {Sun}, \citenamefont {Zhao}, \citenamefont
  {Liu}, \citenamefont {Liu}, \citenamefont {Lu},\ and\ \citenamefont
  {Liu}}]{gu2021}%
  \BibitemOpen
  \bibfield  {author} {\bibinfo {author} {\bibfnamefont {M.}~\bibnamefont
  {Gu}}, \bibinfo {author} {\bibfnamefont {J.}~\bibnamefont {Li}}, \bibinfo
  {author} {\bibfnamefont {H.}~\bibnamefont {Sun}}, \bibinfo {author}
  {\bibfnamefont {Y.}~\bibnamefont {Zhao}}, \bibinfo {author} {\bibfnamefont
  {C.}~\bibnamefont {Liu}}, \bibinfo {author} {\bibfnamefont {J.}~\bibnamefont
  {Liu}}, \bibinfo {author} {\bibfnamefont {H.}~\bibnamefont {Lu}}, \ and\
  \bibinfo {author} {\bibfnamefont {Q.}~\bibnamefont {Liu}},\ }\href {\doibase
  10.1038/s41467-021-23844-z} {\bibfield  {journal} {\bibinfo  {journal} {Nat.
  Commun.}\ }\textbf {\bibinfo {volume} {12}},\ \bibinfo {pages} {3524}
  (\bibinfo {year} {2021})}\BibitemShut {NoStop}%
\bibitem [{\citenamefont {Dai}\ \emph {et~al.}(2022)\citenamefont {Dai},
  \citenamefont {Li}, \citenamefont {Xu}, \citenamefont {Chen},\ and\
  \citenamefont {Xie}}]{dai2022}%
  \BibitemOpen
  \bibfield  {author} {\bibinfo {author} {\bibfnamefont {W.-B.}\ \bibnamefont
  {Dai}}, \bibinfo {author} {\bibfnamefont {H.}~\bibnamefont {Li}}, \bibinfo
  {author} {\bibfnamefont {D.-H.}\ \bibnamefont {Xu}}, \bibinfo {author}
  {\bibfnamefont {C.-Z.}\ \bibnamefont {Chen}}, \ and\ \bibinfo {author}
  {\bibfnamefont {X.~C.}\ \bibnamefont {Xie}},\ }\href {\doibase
  10.1103/PhysRevB.106.245425} {\bibfield  {journal} {\bibinfo  {journal}
  {Phys. Rev. B}\ }\textbf {\bibinfo {volume} {106}},\ \bibinfo {pages}
  {245425} (\bibinfo {year} {2022})}\BibitemShut {NoStop}%
\bibitem [{\citenamefont {Ahn}\ \emph {et~al.}(2022)\citenamefont {Ahn},
  \citenamefont {Xu},\ and\ \citenamefont {Vishwanath}}]{ahn2022}%
  \BibitemOpen
  \bibfield  {author} {\bibinfo {author} {\bibfnamefont {J.}~\bibnamefont
  {Ahn}}, \bibinfo {author} {\bibfnamefont {S.-Y.}\ \bibnamefont {Xu}}, \ and\
  \bibinfo {author} {\bibfnamefont {A.}~\bibnamefont {Vishwanath}},\ }\href
  {\doibase 10.1038/s41467-022-35248-8} {\bibfield  {journal} {\bibinfo
  {journal} {Nat. Commun.}\ }\textbf {\bibinfo {volume} {13}},\ \bibinfo
  {pages} {7615} (\bibinfo {year} {2022})}\BibitemShut {NoStop}%
\bibitem [{\citenamefont {Lei}\ and\ \citenamefont
  {MacDonald}(2023)}]{lei2023}%
  \BibitemOpen
  \bibfield  {author} {\bibinfo {author} {\bibfnamefont {C.}~\bibnamefont
  {Lei}}\ and\ \bibinfo {author} {\bibfnamefont {A.~H.}\ \bibnamefont
  {MacDonald}},\ }\href {https://arxiv.org/abs/2303.14635} {\ ,\ \bibinfo
  {pages} {arXiv:2303.14635} (\bibinfo {year} {2023})}\BibitemShut {NoStop}%
\bibitem [{\citenamefont {Qiu}\ \emph {et~al.}(2023)\citenamefont {Qiu},
  \citenamefont {Tzschaschel}, \citenamefont {Ahn}, \citenamefont {Gao},
  \citenamefont {Li}, \citenamefont {Zhang}, \citenamefont {Ghosh},
  \citenamefont {Hu}, \citenamefont {Wang}, \citenamefont {Liu}, \citenamefont
  {B\'erub\'e}, \citenamefont {Dinh}, \citenamefont {Gong}, \citenamefont
  {Lien}, \citenamefont {Ho}, \citenamefont {Singh}, \citenamefont {Watanabe},
  \citenamefont {Taniguchi}, \citenamefont {Bell}, \citenamefont {Lu},
  \citenamefont {Bansil}, \citenamefont {Lin}, \citenamefont {Chang},
  \citenamefont {Zhou}, \citenamefont {Ma}, \citenamefont {Vishwanath},
  \citenamefont {Ni},\ and\ \citenamefont {Xu}}]{qiu2023}%
  \BibitemOpen
  \bibfield  {author} {\bibinfo {author} {\bibfnamefont {J.-X.}\ \bibnamefont
  {Qiu}}, \bibinfo {author} {\bibfnamefont {C.}~\bibnamefont {Tzschaschel}},
  \bibinfo {author} {\bibfnamefont {J.}~\bibnamefont {Ahn}}, \bibinfo {author}
  {\bibfnamefont {A.}~\bibnamefont {Gao}}, \bibinfo {author} {\bibfnamefont
  {H.}~\bibnamefont {Li}}, \bibinfo {author} {\bibfnamefont {X.-Y.}\
  \bibnamefont {Zhang}}, \bibinfo {author} {\bibfnamefont {B.}~\bibnamefont
  {Ghosh}}, \bibinfo {author} {\bibfnamefont {C.}~\bibnamefont {Hu}}, \bibinfo
  {author} {\bibfnamefont {Y.-X.}\ \bibnamefont {Wang}}, \bibinfo {author}
  {\bibfnamefont {Y.-F.}\ \bibnamefont {Liu}}, \bibinfo {author} {\bibfnamefont
  {D.}~\bibnamefont {B\'erub\'e}}, \bibinfo {author} {\bibfnamefont
  {T.}~\bibnamefont {Dinh}}, \bibinfo {author} {\bibfnamefont {Z.}~\bibnamefont
  {Gong}}, \bibinfo {author} {\bibfnamefont {S.-W.}\ \bibnamefont {Lien}},
  \bibinfo {author} {\bibfnamefont {S.-C.}\ \bibnamefont {Ho}}, \bibinfo
  {author} {\bibfnamefont {B.}~\bibnamefont {Singh}}, \bibinfo {author}
  {\bibfnamefont {K.}~\bibnamefont {Watanabe}}, \bibinfo {author}
  {\bibfnamefont {T.}~\bibnamefont {Taniguchi}}, \bibinfo {author}
  {\bibfnamefont {D.~C.}\ \bibnamefont {Bell}}, \bibinfo {author}
  {\bibfnamefont {H.-Z.}\ \bibnamefont {Lu}}, \bibinfo {author} {\bibfnamefont
  {A.}~\bibnamefont {Bansil}}, \bibinfo {author} {\bibfnamefont
  {H.}~\bibnamefont {Lin}}, \bibinfo {author} {\bibfnamefont {T.-R.}\
  \bibnamefont {Chang}}, \bibinfo {author} {\bibfnamefont {B.~B.}\ \bibnamefont
  {Zhou}}, \bibinfo {author} {\bibfnamefont {Q.}~\bibnamefont {Ma}}, \bibinfo
  {author} {\bibfnamefont {A.}~\bibnamefont {Vishwanath}}, \bibinfo {author}
  {\bibfnamefont {N.}~\bibnamefont {Ni}}, \ and\ \bibinfo {author}
  {\bibfnamefont {S.-Y.}\ \bibnamefont {Xu}},\ }\href {\doibase
  10.1038/s41563-023-01493-5} {\ ,\ \bibinfo {pages} {arXiv:2303.05451}
  (\bibinfo {year} {2023})}\BibitemShut {NoStop}%
\bibitem [{\citenamefont {Nandkishore}\ and\ \citenamefont
  {Levitov}(2011)}]{nandkishore2011}%
  \BibitemOpen
  \bibfield  {author} {\bibinfo {author} {\bibfnamefont {R.}~\bibnamefont
  {Nandkishore}}\ and\ \bibinfo {author} {\bibfnamefont {L.}~\bibnamefont
  {Levitov}},\ }\href {\doibase 10.1103/PhysRevLett.107.097402} {\bibfield
  {journal} {\bibinfo  {journal} {Phys. Rev. Lett.}\ }\textbf {\bibinfo
  {volume} {107}},\ \bibinfo {pages} {097402} (\bibinfo {year}
  {2011})}\BibitemShut {NoStop}%
\bibitem [{\citenamefont {Crassee}\ \emph {et~al.}(2011)\citenamefont
  {Crassee}, \citenamefont {Levallois}, \citenamefont {Walter}, \citenamefont
  {Ostler}, \citenamefont {Bostwick}, \citenamefont {Rotenberg}, \citenamefont
  {Seyller}, \citenamefont {van~der Marel},\ and\ \citenamefont
  {Kuzmenko}}]{crassee2011}%
  \BibitemOpen
  \bibfield  {author} {\bibinfo {author} {\bibfnamefont {I.}~\bibnamefont
  {Crassee}}, \bibinfo {author} {\bibfnamefont {J.}~\bibnamefont {Levallois}},
  \bibinfo {author} {\bibfnamefont {A.~L.}\ \bibnamefont {Walter}}, \bibinfo
  {author} {\bibfnamefont {M.}~\bibnamefont {Ostler}}, \bibinfo {author}
  {\bibfnamefont {A.}~\bibnamefont {Bostwick}}, \bibinfo {author}
  {\bibfnamefont {E.}~\bibnamefont {Rotenberg}}, \bibinfo {author}
  {\bibfnamefont {T.}~\bibnamefont {Seyller}}, \bibinfo {author} {\bibfnamefont
  {D.}~\bibnamefont {van~der Marel}}, \ and\ \bibinfo {author} {\bibfnamefont
  {A.~B.}\ \bibnamefont {Kuzmenko}},\ }\href {\doibase 10.1038/nphys1816}
  {\bibfield  {journal} {\bibinfo  {journal} {Nat. Phys.}\ }\textbf {\bibinfo
  {volume} {7}},\ \bibinfo {pages} {48} (\bibinfo {year} {2011})}\BibitemShut
  {NoStop}%
\bibitem [{\citenamefont {Shimano}\ \emph {et~al.}(2013)\citenamefont
  {Shimano}, \citenamefont {Yumoto}, \citenamefont {Yoo}, \citenamefont
  {Matsunaga}, \citenamefont {Tanabe}, \citenamefont {Hibino}, \citenamefont
  {Morimoto},\ and\ \citenamefont {Aoki}}]{shimano2013}%
  \BibitemOpen
  \bibfield  {author} {\bibinfo {author} {\bibfnamefont {R.}~\bibnamefont
  {Shimano}}, \bibinfo {author} {\bibfnamefont {G.}~\bibnamefont {Yumoto}},
  \bibinfo {author} {\bibfnamefont {J.~Y.}\ \bibnamefont {Yoo}}, \bibinfo
  {author} {\bibfnamefont {R.}~\bibnamefont {Matsunaga}}, \bibinfo {author}
  {\bibfnamefont {S.}~\bibnamefont {Tanabe}}, \bibinfo {author} {\bibfnamefont
  {H.}~\bibnamefont {Hibino}}, \bibinfo {author} {\bibfnamefont
  {T.}~\bibnamefont {Morimoto}}, \ and\ \bibinfo {author} {\bibfnamefont
  {H.}~\bibnamefont {Aoki}},\ }\href {\doibase 10.1038/ncomms2866} {\bibfield
  {journal} {\bibinfo  {journal} {Nat. Commun.}\ }\textbf {\bibinfo {volume}
  {4}},\ \bibinfo {pages} {1841} (\bibinfo {year} {2013})}\BibitemShut
  {NoStop}%
\bibitem [{\citenamefont {Sivadas}\ \emph {et~al.}(2016)\citenamefont
  {Sivadas}, \citenamefont {Okamoto},\ and\ \citenamefont
  {Xiao}}]{sivadas2016}%
  \BibitemOpen
  \bibfield  {author} {\bibinfo {author} {\bibfnamefont {N.}~\bibnamefont
  {Sivadas}}, \bibinfo {author} {\bibfnamefont {S.}~\bibnamefont {Okamoto}}, \
  and\ \bibinfo {author} {\bibfnamefont {D.}~\bibnamefont {Xiao}},\ }\href
  {\doibase 10.1103/PhysRevLett.117.267203} {\bibfield  {journal} {\bibinfo
  {journal} {Phys. Rev. Lett.}\ }\textbf {\bibinfo {volume} {117}},\ \bibinfo
  {pages} {267203} (\bibinfo {year} {2016})}\BibitemShut {NoStop}%
\bibitem [{\citenamefont {Huang}\ \emph {et~al.}(2017)\citenamefont {Huang},
  \citenamefont {Clark}, \citenamefont {Navarro-Moratalla}, \citenamefont
  {Klein}, \citenamefont {Cheng}, \citenamefont {Seyler}, \citenamefont
  {Zhong}, \citenamefont {Schmidgall}, \citenamefont {McGuire}, \citenamefont
  {Cobden}, \citenamefont {Yao}, \citenamefont {Xiao}, \citenamefont
  {Jarillo-Herrero},\ and\ \citenamefont {Xu}}]{huang2017}%
  \BibitemOpen
  \bibfield  {author} {\bibinfo {author} {\bibfnamefont {B.}~\bibnamefont
  {Huang}}, \bibinfo {author} {\bibfnamefont {G.}~\bibnamefont {Clark}},
  \bibinfo {author} {\bibfnamefont {E.}~\bibnamefont {Navarro-Moratalla}},
  \bibinfo {author} {\bibfnamefont {D.~R.}\ \bibnamefont {Klein}}, \bibinfo
  {author} {\bibfnamefont {R.}~\bibnamefont {Cheng}}, \bibinfo {author}
  {\bibfnamefont {K.~L.}\ \bibnamefont {Seyler}}, \bibinfo {author}
  {\bibfnamefont {D.}~\bibnamefont {Zhong}}, \bibinfo {author} {\bibfnamefont
  {E.}~\bibnamefont {Schmidgall}}, \bibinfo {author} {\bibfnamefont {M.~A.}\
  \bibnamefont {McGuire}}, \bibinfo {author} {\bibfnamefont {D.~H.}\
  \bibnamefont {Cobden}}, \bibinfo {author} {\bibfnamefont {W.}~\bibnamefont
  {Yao}}, \bibinfo {author} {\bibfnamefont {D.}~\bibnamefont {Xiao}}, \bibinfo
  {author} {\bibfnamefont {P.}~\bibnamefont {Jarillo-Herrero}}, \ and\ \bibinfo
  {author} {\bibfnamefont {X.}~\bibnamefont {Xu}},\ }\href {\doibase
  10.1038/nature22391} {\bibfield  {journal} {\bibinfo  {journal} {Nature}\
  }\textbf {\bibinfo {volume} {546}},\ \bibinfo {pages} {270} (\bibinfo {year}
  {2017})}\BibitemShut {NoStop}%
\bibitem [{\citenamefont {Gong}\ \emph {et~al.}(2017)\citenamefont {Gong},
  \citenamefont {Li}, \citenamefont {Li}, \citenamefont {Ji}, \citenamefont
  {Stern}, \citenamefont {Xia}, \citenamefont {Cao}, \citenamefont {Bao},
  \citenamefont {Wang}, \citenamefont {Wang}, \citenamefont {Qiu},
  \citenamefont {Cava}, \citenamefont {Louie}, \citenamefont {Xia},\ and\
  \citenamefont {Zhang}}]{gong2017}%
  \BibitemOpen
  \bibfield  {author} {\bibinfo {author} {\bibfnamefont {C.}~\bibnamefont
  {Gong}}, \bibinfo {author} {\bibfnamefont {L.}~\bibnamefont {Li}}, \bibinfo
  {author} {\bibfnamefont {Z.}~\bibnamefont {Li}}, \bibinfo {author}
  {\bibfnamefont {H.}~\bibnamefont {Ji}}, \bibinfo {author} {\bibfnamefont
  {A.}~\bibnamefont {Stern}}, \bibinfo {author} {\bibfnamefont
  {Y.}~\bibnamefont {Xia}}, \bibinfo {author} {\bibfnamefont {T.}~\bibnamefont
  {Cao}}, \bibinfo {author} {\bibfnamefont {W.}~\bibnamefont {Bao}}, \bibinfo
  {author} {\bibfnamefont {C.}~\bibnamefont {Wang}}, \bibinfo {author}
  {\bibfnamefont {Y.}~\bibnamefont {Wang}}, \bibinfo {author} {\bibfnamefont
  {Z.~Q.}\ \bibnamefont {Qiu}}, \bibinfo {author} {\bibfnamefont {R.~J.}\
  \bibnamefont {Cava}}, \bibinfo {author} {\bibfnamefont {S.~G.}\ \bibnamefont
  {Louie}}, \bibinfo {author} {\bibfnamefont {J.}~\bibnamefont {Xia}}, \ and\
  \bibinfo {author} {\bibfnamefont {X.}~\bibnamefont {Zhang}},\ }\href
  {\doibase 10.1038/nature22060} {\bibfield  {journal} {\bibinfo  {journal}
  {Nature}\ }\textbf {\bibinfo {volume} {546}},\ \bibinfo {pages} {265}
  (\bibinfo {year} {2017})}\BibitemShut {NoStop}%
\bibitem [{\citenamefont {Freiser}(1968)}]{freiser1968}%
  \BibitemOpen
  \bibfield  {author} {\bibinfo {author} {\bibnamefont {Freiser}},\ }\href
  {\doibase 10.1109/TMAG.1968.1066210} {\bibfield  {journal} {\bibinfo
  {journal} {IEEE Trans. Magn.}\ }\textbf {\bibinfo {volume} {4}},\ \bibinfo
  {pages} {152} (\bibinfo {year} {1968})}\BibitemShut {NoStop}%
\bibitem [{\citenamefont {Tse}\ and\ \citenamefont
  {MacDonald}(2010)}]{tse2010}%
  \BibitemOpen
  \bibfield  {author} {\bibinfo {author} {\bibfnamefont {W.-K.}\ \bibnamefont
  {Tse}}\ and\ \bibinfo {author} {\bibfnamefont {A.~H.}\ \bibnamefont
  {MacDonald}},\ }\href {\doibase 10.1103/PhysRevLett.105.057401} {\bibfield
  {journal} {\bibinfo  {journal} {Phys. Rev. Lett.}\ }\textbf {\bibinfo
  {volume} {105}},\ \bibinfo {pages} {057401} (\bibinfo {year}
  {2010})}\BibitemShut {NoStop}%
\bibitem [{\citenamefont {Gorbar}\ \emph {et~al.}(2012)\citenamefont {Gorbar},
  \citenamefont {Gusynin}, \citenamefont {Kuzmenko},\ and\ \citenamefont
  {Sharapov}}]{gorbar2012}%
  \BibitemOpen
  \bibfield  {author} {\bibinfo {author} {\bibfnamefont {E.~V.}\ \bibnamefont
  {Gorbar}}, \bibinfo {author} {\bibfnamefont {V.~P.}\ \bibnamefont {Gusynin}},
  \bibinfo {author} {\bibfnamefont {A.~B.}\ \bibnamefont {Kuzmenko}}, \ and\
  \bibinfo {author} {\bibfnamefont {S.~G.}\ \bibnamefont {Sharapov}},\ }\href
  {\doibase 10.1103/PhysRevB.86.075414} {\bibfield  {journal} {\bibinfo
  {journal} {Phys. Rev. B}\ }\textbf {\bibinfo {volume} {86}},\ \bibinfo
  {pages} {075414} (\bibinfo {year} {2012})}\BibitemShut {NoStop}%
\bibitem [{\citenamefont {Tse}\ and\ \citenamefont
  {MacDonald}(2011)}]{tse2011}%
  \BibitemOpen
  \bibfield  {author} {\bibinfo {author} {\bibfnamefont {W.-K.}\ \bibnamefont
  {Tse}}\ and\ \bibinfo {author} {\bibfnamefont {A.~H.}\ \bibnamefont
  {MacDonald}},\ }\href {\doibase 10.1103/PhysRevB.84.205327} {\bibfield
  {journal} {\bibinfo  {journal} {Phys. Rev. B}\ }\textbf {\bibinfo {volume}
  {84}},\ \bibinfo {pages} {205327} (\bibinfo {year} {2011})}\BibitemShut
  {NoStop}%
\bibitem [{\citenamefont {G.}\ \emph {et~al.}(2020)\citenamefont {G.},
  \citenamefont {Peres},\ and\ \citenamefont
  {Fernandez-Rossier}}]{catarina2020}%
  \BibitemOpen
  \bibfield  {author} {\bibinfo {author} {\bibfnamefont {C.}~\bibnamefont
  {G.}}, \bibinfo {author} {\bibfnamefont {N.~M.}\ \bibnamefont {Peres}}, \
  and\ \bibinfo {author} {\bibfnamefont {J.}~\bibnamefont
  {Fernandez-Rossier}},\ }\href {\doibase 10.1088/2053-1583/ab6781} {\bibfield
  {journal} {\bibinfo  {journal} {2D Mater.}\ }\textbf {\bibinfo {volume}
  {7}},\ \bibinfo {pages} {025011} (\bibinfo {year} {2020})}\BibitemShut
  {NoStop}%
\bibitem [{\citenamefont {Kahn}\ \emph {et~al.}(1969)\citenamefont {Kahn},
  \citenamefont {Pershan},\ and\ \citenamefont {Remeika}}]{kahn1969}%
  \BibitemOpen
  \bibfield  {author} {\bibinfo {author} {\bibfnamefont {F.~J.}\ \bibnamefont
  {Kahn}}, \bibinfo {author} {\bibfnamefont {P.~S.}\ \bibnamefont {Pershan}}, \
  and\ \bibinfo {author} {\bibfnamefont {J.~P.}\ \bibnamefont {Remeika}},\
  }\href {\doibase 10.1103/PhysRev.186.891} {\bibfield  {journal} {\bibinfo
  {journal} {Phys. Rev.}\ }\textbf {\bibinfo {volume} {186}},\ \bibinfo {pages}
  {891} (\bibinfo {year} {1969})}\BibitemShut {NoStop}%
\bibitem [{mah()}]{mahan2000}%
  \BibitemOpen
  \href@noop {} {}\bibinfo {note} {Mahan G D 2000 \emph{Many-Particle Physics}
  3rd edn.(New York: Springer) p.507}\BibitemShut {NoStop}%
\bibitem [{giu()}]{giuliani2005}%
  \BibitemOpen
  \href@noop {} {}\bibinfo {note} {Giuliani G and Vignale G 2005 \emph{Quantum
  Theory of the Electron Liquid} 1st edn.(Cambridge University Press)
  p.128}\BibitemShut {NoStop}%
\bibitem [{\citenamefont {Zhou}\ \emph {et~al.}(2020)\citenamefont {Zhou},
  \citenamefont {Zhang},\ and\ \citenamefont {Li}}]{zhou2020}%
  \BibitemOpen
  \bibfield  {author} {\bibinfo {author} {\bibfnamefont {J.}~\bibnamefont
  {Zhou}}, \bibinfo {author} {\bibfnamefont {S.}~\bibnamefont {Zhang}}, \ and\
  \bibinfo {author} {\bibfnamefont {J.}~\bibnamefont {Li}},\ }\href {\doibase
  10.1038/s41427-019-0188-9} {\bibfield  {journal} {\bibinfo  {journal} {NPG
  Asia Materials}\ }\textbf {\bibinfo {volume} {12}},\ \bibinfo {pages} {2}
  (\bibinfo {year} {2020})}\BibitemShut {NoStop}%
\bibitem [{\citenamefont {Cao}\ \emph {et~al.}(2023)\citenamefont {Cao},
  \citenamefont {Jiang}, \citenamefont {Li}, \citenamefont {Tu}, \citenamefont
  {Zhou}, \citenamefont {Zhou},\ and\ \citenamefont {Yao}}]{cao2023}%
  \BibitemOpen
  \bibfield  {author} {\bibinfo {author} {\bibfnamefont {J.}~\bibnamefont
  {Cao}}, \bibinfo {author} {\bibfnamefont {W.}~\bibnamefont {Jiang}}, \bibinfo
  {author} {\bibfnamefont {X.-P.}\ \bibnamefont {Li}}, \bibinfo {author}
  {\bibfnamefont {D.}~\bibnamefont {Tu}}, \bibinfo {author} {\bibfnamefont
  {J.}~\bibnamefont {Zhou}}, \bibinfo {author} {\bibfnamefont {J.}~\bibnamefont
  {Zhou}}, \ and\ \bibinfo {author} {\bibfnamefont {Y.}~\bibnamefont {Yao}},\
  }\href {\doibase 10.1103/PhysRevLett.130.166702} {\bibfield  {journal}
  {\bibinfo  {journal} {Phys. Rev. Lett.}\ }\textbf {\bibinfo {volume} {130}},\
  \bibinfo {pages} {166702} (\bibinfo {year} {2023})}\BibitemShut {NoStop}%
\end{thebibliography}

%

\end{document}